\documentclass[aps,prb,10pt,twocolumn,showpacs,preprintnumbers,amsmath,amssymb,superscriptaddress,longbibliography,floatfix]{revtex4-2}

\usepackage{amsmath}    %
\usepackage{amssymb,environ}
\DeclareMathOperator{\arcsinh}{arcsinh}
\usepackage{graphicx}   %
\usepackage{verbatim}   %
\usepackage{color}      %

\usepackage[normalem]{ulem}

\usepackage{natbib}
\usepackage{enumitem}
\usepackage{multirow}

\usepackage{epsfig}
\usepackage{textcomp}
\usepackage{array}
\usepackage{color}

\usepackage{units}
\usepackage{physics}
\usepackage{dsfont}
\usepackage{hyperref}   %

\usepackage{cleveref}

\crefname{equation}{eq.}{eqs.}
\Crefname{equation}{Eq.}{Eqs.}

\renewcommand{\dd}{\mathrm{d}}
\newcommand{\up}{\uparrow}
\newcommand{\dn}{\downarrow}

\begin{document}

\newcommand{\halfbox}{
  \raisebox{0.5ex}{\rule{0.4pt}{0.7ex}}
  \raisebox{0.3ex}{\rule{1.5ex}{0.4pt}}
  \raisebox{0.5ex}{\rule{0.4pt}{0.7ex}}
}
\setlength{\jot}{2mm}

\title{Quasiparticle-resolved variational theory of Andreev spin qubits}

\author{Teodor Ili\v cin}
\affiliation{Jo\v zef Stefan Institute, Jamova 39, SI-1000 Ljubljana, Slovenia}
\affiliation{Faculty of Mathematics and Physics, University of Ljubljana, Jadranska 19, SI-1000 Ljubljana, Slovenia}

\author{Rok \v Zitko}
\affiliation{Jo\v zef Stefan Institute, Jamova 39, SI-1000 Ljubljana, Slovenia}
\affiliation{Faculty of Mathematics and Physics, University of Ljubljana, Jadranska 19, SI-1000 Ljubljana, Slovenia}

\date{October 12, 2026}

\begin{abstract}
An Andreev spin qubit stores a single unpaired spin, yet its coupling to the superconducting
phase can be dominated by coherent virtual quasiparticle pairs.
We develop a continuum variational description of the odd-parity doublet in a quantum-dot
Josephson junction with Coulomb interaction and spin-dependent background tunneling. Retaining
configurations with up to two Bogoliubov quasiparticles provides compact analytical expressions for energies
and wavefunctions. We benchmark the results against numerical renormalization group calculations
and identify sources of error. We obtain closed-form expressions
for the conventional and spin-dependent Josephson couplings. Despite their small probability,
two-quasiparticle configurations, through coherence with the zero-quasiparticle component, supply exactly half of the
leading spin-dependent Josephson coupling even without Coulomb repulsion and dominate it as the repulsion grows.
Spin-orbit-induced spin transfer between the dot and the leads produces a Josephson-current contribution that is common to both qubit states and finite at zero phase bias when the dot and lead Zeeman energies differ. We also map the localization of the virtual quasiparticles and the spin density in the leads.
The results connect strongly correlated quantum-dot
models to effective Hamiltonians used for Andreev-spin-qubit circuits and clarify how the
wavefunction structure determines their couplings.
\end{abstract}

\maketitle

\section{Introduction}

An Andreev spin qubit (ASQ) encodes quantum information in the spin of a single quasiparticle confined in an Andreev bound state of a superconducting point contact or a hybrid semiconductor-superconductor heterostructure~\cite{chtchelkatchev2003,padurariu2010,Hays2020,hays2021,bargerbos2023}. This hybrid qubit combines the complementary advantages of semiconductor spin qubits and superconducting-circuit platforms: compact devices and electrostatic tunability on the one hand, and circuit-QED readout and long-range circuit-mediated coupling on the other~\cite{pitavidal2023,pitavidal2025}. The essential physics of an ASQ can be captured by the simple phenomenological Hamiltonian
\begin{equation}
    H = E_J \cos\phi + E_S \hat{\mathbf{n}} \cdot \boldsymbol{\sigma} \sin \phi,
    \label{eq:phenom}
\end{equation}
where $\phi$ is the superconducting phase difference, $E_J$ denotes the conventional Josephson energy~\cite{josephson1962}, $E_S$ the spin-dependent Josephson energy~\cite{padurariu2010,zazunov2009,Brunetti2013}, and $\hat{\mathbf{n}}$ the direction of the effective spin-orbit-coupling (SOC) field. The vector of Pauli matrices $\boldsymbol{\sigma}$ acts on the qubit spin degree of freedom. The coupling between the localized spin and the superconducting condensate is modeled by the spin-dependent part of the Hamiltonian~\cite{DellAnna2007}.
This effective description accounts for experimental results on ASQ junctions in most parameter regimes~\cite{bargerbos2023,pitavidal2023} and has been used to model superconducting circuits incorporating such elements~\cite{Kurilovich2026,ipa}, even in setups with long-range coupling between two ASQs~\cite{PitaVidal2024}.
A nonzero $E_S$ generally requires the constriction to involve more than a single orbital~\cite{beri2008,reynoso2012,bargerbos2023}. One microscopic approach is therefore to employ an explicit multiorbital impurity Hamiltonian. In Ref.~\cite{padurariu2010}, the coefficients $E_S$ and $E_J$ were obtained by fourth-order perturbation theory in hopping for a microscopic model of an interacting multilevel quantum dot.
In finite-length Rashba nanowire junctions, mixing with higher-energy transverse subbands likewise generates phase-dependent Andreev spin splitting~\cite{Park2017,Tosi2019SpinOrbit}. A simpler alternative is to assume that only one active orbital lies close to the Fermi level, while the remaining high-energy orbitals are integrated out~\cite{bargerbos2023,Pavesic2024transmon,ilicin2026}. Their effect is then represented by an effective direct interlead hopping term, which accounts for tunneling processes through all inactive orbitals. Josephson junctions combining a dot with direct interlead tunneling have also been studied as interacting interferometers, exhibiting Fano interference and modified singlet--doublet phase boundaries~\cite{Osawa2008,Karrasch2009,Zalom2026}. A related model combining a resonant localized state with a direct-transmission channel has been used to explain measured current--phase relations in InSb--Al nanowire Josephson junctions~\cite{Levajac2024Supercurrent}.

Microwave spectroscopy has resolved both interaction- and SOC-induced structure in the subgap spectrum, including exchange-split two-quasiparticle excitations and spin-orbit-induced mixing
\cite{MatuteCanadas2022,Wesdorp2024}.
In Anderson-type impurity models, confinement-induced interaction effects are described by an on-site Hubbard term for the active orbital, while SOC is incorporated through the hybridization terms. This minimal model suffices to address three central issues: (i) the microscopic origin of the spin splitting and the role of quasiparticles therein; (ii) a quantitative and predictive description of the spin-dependent coupling; and (iii) the role of electron-electron interactions.
Open questions concerning the role of continuum states include the spatial extent of the spin-carrying bound state, as well as the onset of partial Kondo screening with increasing hybridization and the consequences of the resulting spin compensation~\cite{kirsanskas2015,meden2019,Moca2021Kondo,pavesic2024,ilicin2025}.

In previous work, we developed a continuum variational description of the superconducting Anderson model with a single bath~\cite{ilicin2025} and analyzed the even-parity states of quantum-dot Josephson junctions with SOC relevant to Andreev pair qubits~\cite{ilicin2026}. Here we address the complementary odd-parity sector relevant to ASQs. We construct explicit variational wavefunctions containing up to two Bogoliubov quasiparticles (QPs) in the superconducting leads. Superconducting surrogate models approximate the reservoir hybridization by fitting a small number of effective levels~\cite{Baran2023}, while the chain-expansion (ChE) approach constructs finite superconducting chains from Pad\'e approximants of the hybridization self-energy, without fitting~\cite{Bobok2025}. Our approach is complementary: we retain the continuum and truncate the number of Bogoliubov excitations. We first solve the phase-biased junction without direct interlead hopping and subsequently include it perturbatively.

This construction provides compact expressions for the Josephson couplings at a finite
superconducting gap, together with the corresponding QP composition. In the weak-hybridization
limit, we derive an analytical expression for $E_S$. We show why a 2-QP component with a small probability can nevertheless be essential for the magnitude of the spin splitting: the tunneling operator probes coherence between sectors as well as their probabilities (a sector specifies the number and configuration of QPs in the leads). The resulting description connects the interacting microscopic model to the effective $E_J$--$E_S$ Hamiltonians used for circuits incorporating ASQs.

We benchmark the variational solution against numerical renormalization group (NRG) and density-matrix renormalization group (DMRG) results. We also test convergence by numerically finding the variational solution with larger QP cutoffs. We perform the numerical variational and DMRG calculations with our open-source package \texttt{qdjj\_solver}~\cite{BCSQPSolver2026}, controlling bath resolution and many-body convergence independently.

The manuscript is organized as follows: we specify the microscopic model, construct and benchmark the variational state, evaluate the Josephson couplings and compare our results with those of previous approaches. We relate SOC-induced spin redistribution to the magnetic Josephson response before discussing QP localization and the spatial spin distribution and summarizing our main results. Supporting derivations are given in the appendices and the Supplemental Material~\cite{asq_supplemental}.
    
\section{Model}

\begin{figure*}[t]
    \centering
    \includegraphics{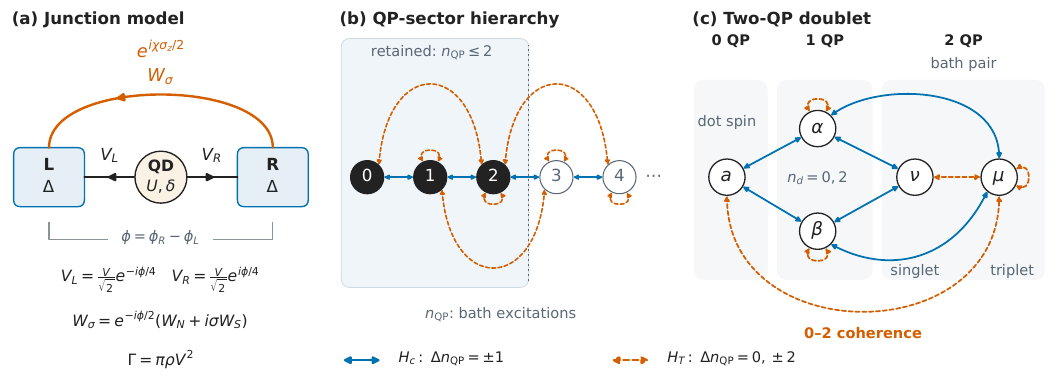}
    \caption{Microscopic two-path model and QP-sector couplings of an Andreev spin qubit.
    (a) A single interacting quantum-dot orbital (QD), with Coulomb
    repulsion $U$ and detuning $\delta$, couples to two superconducting
    leads with gap $\Delta$ via spin-independent amplitudes $V_L,V_R$.
    The curved path represents effective tunneling through inactive
    orbitals, with spin-dependent amplitude $W_\sigma$. The factor $e^{i\chi\sigma_z/2}$
    denotes its spin-dependent phase.
    Arrows in (a) specify the coefficients of $c^\dagger_{lk\sigma}d_\sigma$
    and $c^\dagger_{Lk\sigma}c_{Rk'\sigma}$, whereas the conjugate pairs are implicit.
    (b) Bath-QP sectors relative to the isolated BCS leads. Dot--lead hopping
    $H_c$ connects adjacent sectors (blue solid links), while $H_T$ scatters
    within sectors or creates/annihilates QP pairs (vermilion dashed links).
    Shading marks the space $n_{\mathrm{QP}}\leq2$ that we retain for
    analytical derivations in this work.
    (c) Components of the $S_z = 1/2$ doublet Ansatz at $W=0$ and $\delta=0$; the variational wavefunction contains both bath singlet ($\nu$) and triplet pairs ($\mu$).
    Dashed links here represent the spin-dependent part of $H_T$ projected onto this Ansatz, including the bare impurity--triplet ($a$--$\mu$) coherence. }
    \label{fig:setup}
\end{figure*}

We model the quantum dot as a single interacting impurity orbital coupled to two superconducting leads, using the Anderson impurity model (AIM)~\cite{anderson1961,meden2019,kirsanskas2015,martinrodero2011}. The two-path setup, including the effective direct interlead tunneling, is shown in Fig.~\ref{fig:setup}(a). The Hamiltonian is
\begin{equation}
    H = H_{\mathrm{imp}} + \sum_l H_{\mathrm{BCS},l} + \sum_l H_{c, l} + H_T \>.
\end{equation}

The impurity part is
\begin{equation}
    H_\mathrm{imp} = \frac{U}{2} \left( n- \nu \right) ^2 = \frac{U}{2} (n-1)^2 - U \delta (n-1) + \text{const.}
\end{equation}
Here, $U$ sets the strength of the Coulomb interaction, while the deviation from the
particle-hole-symmetric point is parametrized by $\delta =\nu -1$. In analytical calculations in
this work, we consider $\delta=0$. The impurity occupancy operator is defined as $n=\sum_\sigma d^\dag_\sigma d_\sigma$, where $d_{\sigma}$ is the annihilation operator for the impurity orbital with spin $\sigma$.

The index $l\in \{L, R\}$ labels the leads. We assume $\Delta_{L}=\Delta_{R} = \Delta$ and choose a gauge in which the superconducting phases are encoded in the complex hopping amplitudes. Both leads are described by the Bardeen--Cooper--Schrieffer (BCS) Hamiltonian
\begin{equation}
    H_{\mathrm{BCS},l} = \sum_{k\sigma} \epsilon_k c^\dag_{lk\sigma} c_{lk\sigma}
    -\left( \Delta \sum_k c^\dag_{lk\uparrow} c^\dag_{l,-k\downarrow} + \text{H.c.} \right)
\end{equation}
with $\Delta>0$. Here $\epsilon_k$ is the normal-state single-particle energy measured from the chemical potential. The operator $c_{lk\sigma}^\dag$ creates an electron in mode $k \sigma$ in lead $l$. The corresponding BCS wavefunctions are
\begin{equation}
    \ket{\Phi_{l}} = \prod_k \left( u_{k} + v_{k} c^\dag_{lk\uparrow} c^\dag_{l,-k,\downarrow} \right) \ket{0}.
    \label{eq:BCS}
\end{equation}
In the chosen gauge, the coherence factors are the same for both leads:
\begin{equation}
    u_{k} = \sqrt{\frac{1}{2} \left( 1 + \frac{\epsilon_k}{E_k} \right)}, \quad
    v_{k} = \sqrt{\frac{1}{2} \left( 1 - \frac{\epsilon_k}{E_k} \right)},
\end{equation}
with $E_k = \sqrt{\epsilon_k^2 + \Delta^2}$. The QP operators are likewise defined such that they do not depend on the phase~\cite{bcs,tinkham}:
\begin{align}
 \gamma_{lk\uparrow} &= u_{k} c_{lk\uparrow} - v_{k} c^\dag_{l,-k\downarrow}, \\
 \gamma^\dag_{l,-k\downarrow} &= v_{k} c_{lk\uparrow} + u_{k} c^\dag_{l,-k\downarrow}.
\end{align}
We include only the spin-conserving (scalar) terms in the coupling between the leads and the dot:
\begin{equation}
    H_{c, l} = \frac{1}{\sqrt{N}} \sum_{k\sigma} V_{l,k} c^\dag_{l, k, \sigma} d_{\sigma} + \mathrm{H.c.} \>.
\end{equation}
Here $N$ is the number of orbital modes $k$ in each lead. We assume $k$-independent hopping, $V_{l,k} = V_l$, and symmetric amplitudes:
\begin{equation}
    |V_{L}| = |V_{R}| = \frac{|V|}{\sqrt{2}} \>.
\end{equation}
Both leads have the same constant normal-state density of states per spin, $\rho=1/(2D)$, with bandwidth $2D$. We use $D/\Delta=100$ in numerical calculations.
The dot--lead coupling is fully encoded in the energy dependence of the normal-state hybridization function $\Gamma(\epsilon)$~\cite{Bulla1997Anderson}, here taken to be flat within the band (a boxcar function).
We define the total hybridization strength as $\Gamma = \pi\rho\bigl(|V_L|^2+|V_R|^2\bigr) = \pi \rho |V|^2$.
We work in the gauge with $-\phi_L = \phi_R = \phi/2$ and $V_{l} = \frac{V}{\sqrt{2}} e^{i \phi_l / 2}$, taking $V=|V|>0$.

To model an ASQ, we extend the standard AIM by including the direct-coupling term $H_T$. In this setup, spin splitting occurs only when the two tunneling paths (through the active orbital and through the high-energy orbitals represented by $H_T$) have different effective SOC, i.e., different ratios of scalar to spin-dependent tunneling amplitudes~\cite{beri2008,reynoso2012,bargerbos2023}. It is convenient to orient the SOC axis along the $z$ axis and to include all spin-dependent processes in the direct tunneling term:
\begin{equation}
    H_T = \frac{1}{N} \sum_{k, k', \sigma} W_{\sigma} c_{L, k, \sigma}^\dag c_{R, k', \sigma} + \mathrm{H.c.} \>,
    \label{eq:H_T}
\end{equation}
where
\begin{equation}
    \begin{split}
        W_{\sigma} &= W e^{- i ( \phi - \sigma \chi) / 2} = e^{-i\phi/2}(W_N+i\sigma W_S) \\
        W &= \sqrt{W_N^2 + W_S^2} \\
        W_N &= W\cos(\chi/2), \quad W_S=W\sin(\chi/2) \>.
    \end{split}
\end{equation}
Here $W_N$ and $W_S$ represent the scalar and spin-dependent tunneling amplitudes, respectively, and $\sigma=+1$/$-1$ for spin $\up$/$\dn$. The angle $\chi$ parametrizes the relative spin rotation between the two tunneling paths. No generality is lost by choosing this representation for the two-path model considered here.

In the next section, we first construct a variational solution for $W=0$, where the equations
simplify considerably due to symmetry.

\section{Variational solution at zero background tunneling}
\label{sec:ph-symmetric}
\subsection{Symmetry properties}

For $W=0$ and $\delta=0$, particle-hole conjugation and the mirror exchange $L\leftrightarrow R$ act on $H$ by mapping $\phi\to-\phi$. 
Hence, both leave $H$ invariant only when composed with complex conjugation, i.e., as antiunitary operators. 
While antiunitary symmetries can constrain the eigenstates, they supply no conserved quantum numbers and do not decompose $H$ into a block structure. 
However, the product of the two symmetries is unitary and commutes with $H$ for all $\phi$~\cite{ilicin2026}:
\begin{equation}
    \eta : = \begin{cases}
        d_{\sigma} \rightarrow - d_{\bar{\sigma}}^\dag \\
        c_{l,k,\sigma} \rightarrow c^\dag_{\bar{l}, \bar{k}, \bar{\sigma}}
    \end{cases} \>.
\end{equation}
Here, bars swap the lead or spin index, while $\bar k$ denotes particle-hole reflection of the energy label, such that $\bar{\bar k}=k$, $\overline{-k}=-\bar k$. 
For the symmetry to hold, we assume $\epsilon_{\bar k}=-\epsilon_k$ and $\epsilon_{-k}=\epsilon_k$, so that $u_{\bar k}=v_k$ and $v_{\bar k}=u_k$. 
Since the Bogoliubov transformation pairs $k$ with $-k$, the QP operators transform as
\begin{equation}
    \gamma_{l, k, \sigma} \rightarrow -\sigma \gamma_{\bar{l}, -\bar{k}, \sigma} \>.
\end{equation}
As $H$ commutes with $\eta$, the Hamiltonian is block-diagonal in a basis of states with definite $\eta$-parity. We use this to simplify both the construction of the variational Ansatz and the subsequent minimization procedure by introducing two QP channels:
\begin{equation}
    \gamma_{s, k, \sigma} = \frac{1}{\sqrt{2}} \left( \gamma_{L, k, \sigma} + s\gamma_{R, -\bar{k}, \sigma} \right) \>,
    \label{eq:qp_channels}
\end{equation}
where the channel index is $s=\pm 1$. The $+$ and $-$ channel labels are distinct from $\eta$-parity, which also depends on spin. Under $\eta$, the channel operators transform as
\begin{equation}
     \gamma_{s, k, \sigma} \rightarrow - {\sigma} s \gamma_{s, k, \sigma} \>.
\end{equation}
Thus $\gamma_{s,k,\sigma}$ has $\eta$-parity $-\sigma s$: for spin up the $-$ channel is even and the $+$ channel odd, with the assignment reversed for spin down.
This spin dependence reflects that while $\eta$ commutes with the total spin of the dot and leads, $S^2_{\mathrm{tot}}$, and its $z$ component $S^z_{\mathrm{tot}}$, it anticommutes with $S^{\pm}_{\mathrm{tot}}$, so the two members of a spin doublet have opposite $\eta$-parity.
The Hamiltonian is not explicitly $2\pi$-periodic, with $H(\phi + 2\pi)$ and $H(\phi)$ related by a gauge transformation. If this gauge structure is ignored, the $+$ and $-$ channels appear to swap under $\phi \to \phi + 2\pi$, even though the Bogoliubov basis does not explicitly depend on $\phi$. This produces a spurious $4\pi$ periodicity, which disappears once the appropriate gauge transformation is taken into account (see Sec.~{I} of the Supplemental Material).
We build the Ansatz from
\begin{equation}
    \ket{\Phi_0} = \ket{\halfbox} \otimes \ket{\Phi_L} \otimes \ket{\Phi_R} \>,
\end{equation}
where the first ket denotes the empty impurity orbital, while the latter two are the BCS wavefunctions defined in Eq.~\eqref{eq:BCS}. We choose the many-body phase of $\eta$ so that the lead part of $\ket{\Phi_0}$ is invariant and $\eta\ket{\Phi_0}=d_\up^\dag d_\dn^\dag\ket{\Phi_0}$. Thus $\eta$ maps the empty impurity orbital onto the doubly occupied state and vice versa, with $\eta^2=1$. Consequently, impurity states of definite $\eta$-parity transform as follows:
\begin{align}
   d^{\dag}_\sigma \ket{\halfbox} &\rightarrow \sigma d^{\dag}_\sigma \ket{\halfbox}  \\
    \frac{1 \pm d^{\dag}_\up d^{\dag}_\dn}{\sqrt{2}} \ket{\halfbox}  &\rightarrow \pm \frac{1 \pm d^{\dag}_\up d^{\dag}_\dn}{\sqrt{2}} \ket{\halfbox}   \>.
\end{align}

A nonzero scalar background amplitude $W_N$ breaks the $\eta$ symmetry, whereas the purely spin-dependent term preserves it: $\eta H_T(W_N,W_S)\eta^{-1}=H_T(-W_N,W_S)$. 
To retain a compact finite-hybridization description of the leading background-tunneling effects,
we first solve the $W=0$ problem variationally and then treat $H_T$ perturbatively. Nonperturbative inclusion of $H_T$ requires opposite-$\eta$ mixing for
the $W_N$ term, doublet--quartet mixing for the $W_S$ term, and the resummation of quasiparticle scattering within the retained sectors.

The numerical variational expansion solver can handle general situations where the even and odd $\eta$-sectors mix, for
instance when $\delta \neq 0$ or $W_N \neq 0$ (see Appendix~\ref{newappA} for details).

\subsection{Variational equations}

At $W=0$, the Hamiltonian conserves both $S_{\mathrm{tot}}^2$ and $S^z_{\mathrm{tot}}$, and common eigenstates of these operators and $\eta$ may be chosen. We restrict the discussion to the $S_z=1/2$, $\eta$-even branch connected to the bare spin-up impurity. Its spin-rotation partner is degenerate in energy and is defined by $\ket{\Psi_\dn}=S^-_{\mathrm{tot}}\ket{\Psi_\up}$.\footnote{The lowering operator commutes with the singlet charge factor $d_\up^\dag d_\dn^\dag$. Its action is not a literal interchange of every displayed spin index or a complex conjugation of the amplitudes. Ordinary time reversal relates opposite phase biases, while the finite-phase degeneracy here follows from spin-rotation symmetry.}

For weak and intermediate hybridization ($\Gamma\lesssim\Delta$), we approximate the lowest odd-parity doublet using the variational Ansatz $\ket{\Psi}=\sum_i\ket{\psi_i}$, retaining configurations with up to two bath QPs~\cite{ilicin2025,ilicin2026,rozhkov2000}. Its accuracy at finite hybridization depends on the observable, and we assess it by comparing the results against numerical benchmarks below. 
At $W=0$, the 1-QP truncation already captures the leading weak-hybridization energy shift, but remains phase independent: the conventional Josephson response first appears when virtual QP pairs are retained. In the $\eta$-even sector, the components of $\ket{\Psi}$ are
\begin{equation}
        \ket{\psi_1} = a d_{\up}^{\dag} \ket{\Phi_0} 
            \label{eq:ansatz}
\end{equation}
\begin{equation}
        \ket{\psi_2} = \sum_k \frac{\alpha_k}{\sqrt{2 N}} \left(1 + d_{\up}^\dag d_{\dn}^\dag \right) \gamma_{-, k, \up}^\dag \ket{\Phi_0} 
\end{equation}
\begin{equation}        
        \ket{\psi_3} = \sum_k \frac{\beta_k}{\sqrt{2 N}} \left(1 - d_{\up}^\dag d_{\dn}^\dag \right) \gamma_{+, k, \up}^\dag \ket{\Phi_0}
\end{equation}
\begin{equation}
        \ket{\psi_4} = \sum_{kk'} \frac{\nu_{kk'}}{N \sqrt{2}} d_{\up}^\dag
                            \left( \gamma_{+, k,\up}^\dag \gamma_{-, k',\dn}^\dag - \gamma_{+, k,\dn}^\dag \gamma_{-, k',\up}^\dag \right) \ket{\Phi_0}
\end{equation}
\begin{equation}
\begin{split}                            
        \ket{\psi_5} &= \sum_{kk'} \frac{\mu_{kk'}}{N \sqrt{6}} \bigg( 2 d_{\dn}^\dag \gamma_{+, k,\up}^\dag \gamma_{-, k', \up}^\dag \\ &\quad \quad \quad - d_{\up}^\dag \left(\gamma_{+, k,\up}^\dag \gamma_{-, k', \dn}^\dag + \gamma_{+, k, \dn}^\dag \gamma_{-, k', \up}^\dag \right) \bigg) \ket{\Phi_0}.
    \end{split}  
    \label{eq:ansatz5}
\end{equation}
The 0-QP component $\ket{\psi_1}$ carries the spin on the singly occupied impurity. This component is also referred to as the vacuum state, which is dressed by the remaining components. The 1-QP components $\ket{\psi_2},\ket{\psi_3}$ describe virtual empty/doubly occupied impurity configurations, with the spin carried by a lead QP. The 2-QP components retain a singly occupied impurity and couple it to a singlet pair ($\nu$) or a triplet pair ($\mu$), in both cases giving total spin $1/2$.

We measure energy ($\epsilon$) relative to the decoupled-lead BCS energy $E_{\mathrm{BCS}}^{(0)}=\sum_{lk}(\epsilon_k-E_k)$. Requiring the functional $f = \expval{H-E_{\mathrm{BCS}}^{(0)}}{\Psi} - \epsilon (\braket{\Psi} - 1)$ to be stationary with respect to the variational parameters $a^*, \alpha_k^*, \beta_k^*, \nu_{kk'}^*, \mu_{kk'}^*$ gives the equations for the lowest odd-parity state within this Ansatz:
\begin{align}
    \epsilon a &= \frac{V}{ N \sqrt{2}} \sum_k a_k^* \alpha_k + \frac{V}{N  \sqrt{2 }} \sum_k s_k^* \beta_k  \label{eq:a} \\
    N_k \alpha_k &= \frac{V}{\sqrt{2}} a a_k - \frac{V}{2 N}  \sum_{k'} s_{k'}^* \left( \sqrt{3} \mu_{k'k} + \nu_{k'k} \right) \label{eq:alpha} \\ 
    N_k \beta_k &= \frac{V}{\sqrt{2}} a s_k - \frac{V}{2 N}  \sum_{k'} a_{k'}^* \left( \sqrt{3} \mu_{kk'} - \nu_{kk'} \right) \label{eq:beta} \\ 
    N_{kk'} \nu_{kk'} &= -\frac{V}{2} s_k \alpha_{k'} + \frac{V}{2} \beta_k a_{k'} \label{eq:nu} \\
    N_{kk'} \mu_{kk'} &= -\frac{\sqrt{3}}{2} V s_k \alpha_{k'} - \frac{\sqrt{3}}{2} V \beta_k a_{k'} \label{eq:mu} \>. 
\end{align}
Here, we have introduced the following complex combinations of the coherence factors:
\begin{equation}
    s_k = e^{-i\phi/4} u_k + e^{i\phi/4} v_k, \quad
    a_k = e^{-i\phi/4} u_k - e^{i\phi/4} v_k \>,
    \label{eq:coherence_combinations}
\end{equation}
and the auxiliary functions:
\begin{equation}
    N_k = \epsilon - \left( E_k + U/2 \right),  \quad 
    N_{kk'} = \epsilon - (E_k + E_{k'} ) \>.
\end{equation}
Next, we solve \Cref{eq:a,eq:nu,eq:mu} for $a$, $\nu_{kk'}$, and $\mu_{kk'}$ and insert the expressions into \Cref{eq:alpha,eq:beta}. In the continuum limit, we replace $\epsilon_k$ by $\omega$ and use $f_k\to f(\omega)$ for each energy-dependent coefficient. The sums become integrals according to $(1/N)\sum_k\bullet\to\rho\int_{-D}^{D}\bullet\,\dd\omega$.
The continuum QP energy and denominators are
\begin{equation}
    \begin{aligned}
        E(\omega)&=\sqrt{\omega^2+\Delta^2},\\
        N(\epsilon,\omega)&=\epsilon-E(\omega)-U/2,\\
        N(\epsilon,\omega,\omega')&=\epsilon-E(\omega)-E(\omega').
    \end{aligned}
    \label{eq:continuum_denominators}
\end{equation}
We suppress the eigenenergy argument $\epsilon$ when it is implicit.
The resulting integral equations are
\begin{align}
    N(\omega) \alpha(\omega) &= \frac{\Gamma a(\omega) }{2\pi \epsilon}   \left[ A + B \right] + \frac{\Gamma \alpha(\omega)}{\pi}  \int \frac{|s(\omega')|^2 \dd \omega'}{N(\omega,\omega')} \notag \\
    & \quad +  \frac{\Gamma a(\omega)}{2 \pi}  \int \frac{s^* (\omega')\beta(\omega')}{N(\omega,\omega')} \dd \omega', \label{eq:alpha_final}
\end{align}
\begin{align}
    N(\omega) \beta(\omega) &= \frac{\Gamma s(\omega)}{2\pi \epsilon}   \left[ A + B \right] + \frac{\Gamma \beta(\omega)}{\pi}  \int \frac{|a(\omega')|^2 \dd \omega'}{N(\omega,\omega')} \notag \\
    & \quad +  \frac{\Gamma s(\omega)}{2 \pi}  \int \frac{a^*(\omega') \alpha(\omega')}{N(\omega,\omega')} \dd \omega',\label{eq:beta_final}
\end{align}
where   $A = \int \alpha(\omega') a^*(\omega') \dd \omega'$, and $B= \int \beta(\omega') s^*(\omega') \dd \omega'$. Unspecified energy integrals run over $[-D,D]$. We are left with two coupled integral equations for $\alpha(\omega)$ and $\beta(\omega)$. 

The coherence combinations satisfy $s(\omega;-\phi)=s^*(\omega;\phi)$ and
$s(\omega;\phi+2\pi)=-ia(\omega;\phi)$, and likewise with $s$ and $a$ interchanged. Hence, with the
0-QP coefficient $a$ real, the solutions of Eqs.~\eqref{eq:a}--\eqref{eq:mu} obey
\begin{equation}
    \begin{aligned}
        \alpha(\omega;2\pi-\phi)&=-i\beta^*(\omega;\phi), \\
        \beta(\omega;2\pi-\phi)&=-i\alpha^*(\omega;\phi), \\
        \nu(\omega,\omega';2\pi-\phi)&=\nu^*(\omega',\omega;\phi),\\
        \mu(\omega,\omega';2\pi-\phi)&=-\mu^*(\omega',\omega;\phi).
    \end{aligned}
    \label{eq:phase_reflection}
\end{equation}
Reflection about $\phi=\pi$ thus exchanges the $+$ and $-$ channels. Channel-resolved quantities,
such as $|\alpha|^2$ and $|\beta|^2$, coincide at $\phi=\pi$ and have period $4\pi$, whereas
channel-summed weights and physical observables are $2\pi$-periodic
(Sec.~{I} of the Supplemental Material).

\subsection{Kernel-truncation schemes}

The coupled integral equations \eqref{eq:alpha_final} and \eqref{eq:beta_final} are exact stationarity conditions within the 2-QP projection. We derive approximations that reduce these equations to a single scalar equation for the energy. After solving Eq.~\eqref{eq:alpha_final} for $\alpha(\omega)$ and Eq.~\eqref{eq:beta_final} for $\beta(\omega)$, we apply one of three levels of approximation:
\begin{enumerate}
    \item We omit the 2-QP part of the Ansatz.
    
    \item We drop the final (coupling) term in each equation.
    
    \item We insert the expressions into one another and neglect feedback terms proportional to $\Gamma^2\alpha(\omega)$ and $\Gamma^2\beta(\omega)$, which first affect the energy at order $\Gamma^3$.
\end{enumerate}    
In each case, we multiply the first equation by $a^*(\omega)$ and the second by $s^*(\omega)$, and integrate both over $\omega$. Adding the equations and canceling the common factor $[A+B]$ gives one implicit equation for $\epsilon$ per approximation level:
\begin{align}
    \epsilon^{(1)} &= \frac{\Gamma}{\pi} \int \frac{\dd \omega}{N(\epsilon^{(1)}, \omega)} \label{eq:1QP}, \\
    \epsilon^{(2)} &= \frac{\Gamma}{2\pi} \left[ \int \frac{|a(\omega)|^2 \dd \omega}{N_s(\epsilon^{(2)}, \omega)} + \int \frac{|s(\omega)|^2 \dd \omega}{N_a(\epsilon^{(2)}, \omega)} \right] \label{eq:2QP_app1}, \\
    \epsilon^{(3)} &= \frac{\Gamma}{2\pi} \bigg[ \int \frac{|a(\omega)|^2 \dd \omega}{N_s(\epsilon^{(3)}, \omega)} + \int \frac{|s(\omega)|^2 \dd \omega}{N_a(\epsilon^{(3)}, \omega)} \nonumber \\ &\qquad + \frac{\Gamma}{\pi} \int \int  \frac{|a(\omega)|^2 |s(\omega')|^2 \dd \omega \dd \omega'}{N_s(\epsilon^{(3)}, \omega)N(\epsilon^{(3)}, \omega,\omega') N_a(\epsilon^{(3)}, \omega')}
    \bigg]  \>,
    \label{eq:2QP_app2}
\end{align}
where the two new auxiliary functions are
\begin{align}
    N_s (\epsilon, \omega) &= N(\epsilon, \omega) - \frac{\Gamma}{\pi} \int \frac{|s(\omega')|^2 \dd \omega'}{N(\epsilon, \omega, \omega')} \>, \\
    N_a (\epsilon, \omega) &= N(\epsilon, \omega) - \frac{\Gamma}{\pi} \int \frac{|a(\omega')|^2 \dd \omega'}{N(\epsilon, \omega, \omega')} \>.
\end{align}
The restored $\epsilon$ arguments in the $N$-functions emphasize that these eigenvalue equations are implicit transcendental equations.

Since the Ansatz without 2-QP components correctly reproduces the regular $\Gamma \to 0$ asymptotic behavior, the expansion to linear order in $\Gamma$ follows directly from Eq.~\eqref{eq:1QP}. For fixed $U/\Delta>-2$, $\Gamma\ll\Delta$, and $D\gg\max(\Delta,|U|)$, we obtain
\begin{equation}
    \begin{split}
            \epsilon^{(\Gamma)} &\approx -\frac{\Gamma}{\pi} G(U) \>, \\
            G(U) &= 2 c_B - \frac{4U}{\sqrt{(2\Delta)^2 - U^2}} \arctan{\sqrt{\frac{2\Delta - U}{2\Delta + U}}}, \\
    \end{split}
    \label{eq:expansion}
\end{equation}
with
\begin{equation}
            c_B = \arcsinh(D/\Delta) .
\end{equation}
The displayed expression for $G(U)$ applies directly for $-2\Delta<U<2\Delta$. For $U>2\Delta$, its real continuation is
\begin{equation}
    G(U)=2c_B-\frac{2U}{\sqrt{U^2-4\Delta^2}}
    \operatorname{arcosh}\!\left(\frac{U}{2\Delta}\right),
\end{equation}
with the continuous value $G(2\Delta)=2c_B-2$.
The linear coefficient diverges at $U = - 2\Delta$. This is a signature of anomalous $\Gamma^{2/3}$-scaling that arises from the atomic-limit ($\Gamma=0$) degeneracy between $\ket{\psi_1}$ and the gap-edge 1-QP states $\ket{\psi_2}$ and $\ket{\psi_3}$. An analogous effect occurs in the even-parity sector at $U=2\Delta$~\cite{ilicin2025}. Near $U=-2\Delta$, the linear expansion is valid only when $\Gamma G(U)/\pi\ll\Delta+U/2$; otherwise, the anomalous behavior takes over, as discussed in Appendix~\ref{appX}. Variational theory correctly captures the nonanalytic $\Gamma^{2/3}$ asymptotics.

The variational method also gives direct access to the structure of the wavefunction. As in the treatment of $\epsilon$, we derive approximate expressions for the coefficients $\alpha(\omega)$ and $\beta(\omega)$ at different approximation levels:
\begin{align}
    \begin{split}
        \alpha^{(1)}(\omega) &= a \frac{V}{\sqrt{2}} \frac{a(\omega)}{N(\epsilon,\omega)} \\
        \beta^{(1)}(\omega) &= a \frac{V}{\sqrt{2}} \frac{s(\omega)}{N(\epsilon, \omega)}
    \end{split} \label{eq:wf1QP} \\[1.4em]
    \begin{split}
        \alpha^{(2,3)}(\omega) &= a \frac{V}{\sqrt{2}} \frac{a(\omega)}{N_s(\epsilon,\omega)} \left[ 1+A^{(2,3)}(\epsilon, \omega) \right]\\
        \beta^{(2,3)}(\omega) &= a \frac{V}{\sqrt{2}} \frac{s(\omega)}{N_a(\epsilon, \omega)} \left[1+B^{(2,3)}(\epsilon, \omega) \right] \>.
    \label{eq:wf2QP}
    \end{split}
\end{align}
Equation~\eqref{eq:wf1QP} gives the exact coefficients within the 1-QP Ansatz, whereas Eq.~\eqref{eq:wf2QP} represents the approximate 2-QP solutions. Here, setting $A^{(2)}(\epsilon, \omega) = B^{(2)}(\epsilon, \omega) = 0$ corresponds to the level-2 approximation, while taking
\begin{equation}
    \begin{split}
        A^{(3)}(\epsilon, \omega) = \frac{\Gamma}{2 \pi} \int \frac{|s(\omega')|^2 \dd \omega'}{N(\epsilon, \omega, \omega') N_a(\epsilon, \omega')} \\ 
        B^{(3)}(\epsilon, \omega) = \frac{\Gamma}{2 \pi} \int \frac{|a(\omega')|^2 \dd \omega'}{N(\epsilon, \omega, \omega') N_s(\epsilon, \omega')} 
    \end{split}
\end{equation}
corresponds to the level-3 approximation. The coefficients $\nu$ and $\mu$ can be expressed in terms of $\alpha$ and $\beta$, and $a$ can be determined from the normalization condition:
\begin{equation}
    1 = |a|^2 + |\alpha|^2 + |\beta|^2 + |\mu|^2 + |\nu|^2 \>.
    \label{eq:wavefunction_normalization}
\end{equation}
Here the integrated component weights are
\begin{equation}
    \begin{aligned}
        |\alpha|^2 &= \rho \int |\alpha(\omega)|^2 \dd \omega,\\
        |\nu|^2 &= \rho^2 \iint |\nu(\omega,\omega')|^2 \dd\omega\,\dd\omega',
    \end{aligned}
    \label{eq:component_weights}
\end{equation}
with analogous definitions for $|\beta|^2$ and $|\mu|^2$. We fix the overall phase by taking $a>0$.
Figure~\ref{fig:wf_omega} in Appendix~\ref{appwf} shows the wavefunction coefficients for a representative parameter set.

\subsection{Energy benchmarks and truncation errors}

\begin{figure}[!htbp]
    \centering
    \includegraphics{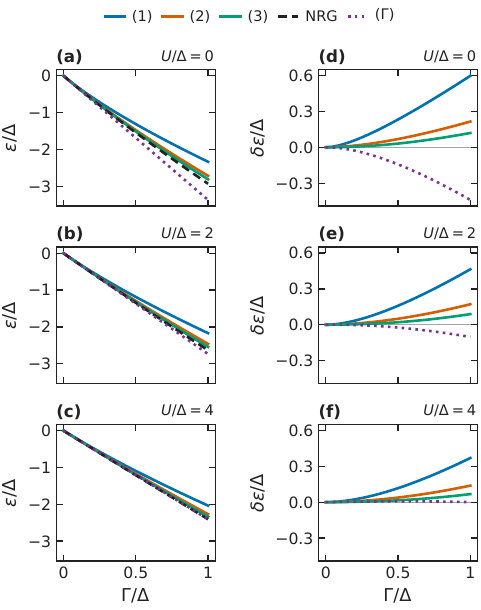}
    \caption{Energy benchmarks. (a)--(c) Doublet energy $\epsilon$ versus hybridization $\Gamma$ at $\phi=0$ for $U/\Delta=0,2,4$, respectively. Solid: variational solutions at different approximation levels; dashed: NRG; dotted: linear expansion $\epsilon^{(\Gamma)}$.
    (d)--(f) Discrepancies $\delta\epsilon^{(i)}=\epsilon^{(i)}-\epsilon^{(\mathrm{NRG})}$ of the three approximation levels and the linear expansion relative to NRG at the corresponding $U$.
    }
    \label{fig:eps}
\end{figure}

The scalar energy equations are solved numerically using Brent's bracketed root-finding method. The energy integrals are evaluated using Gauss--Legendre quadrature after the substitution $\omega=\Delta\sinh t$, with the energy-dependent denominators recomputed at each trial energy. For the energy curves, numerical convergence is checked by refining the quadrature grid and verifying the scalar-equation residual.

In Fig.~\ref{fig:eps}, we compare the solutions at $\phi=0$ with the benchmark NRG results~\cite{wilson1975,bulla2008,yoshioka2000}. NRG details are given in Appendix~\ref{app:nrg}.
Including 2-QP configurations brings the results into better agreement with the reference. The further improvement from $\epsilon^{(2)}$ to $\epsilon^{(3)}$ is much smaller, indicating that the mixed kernel contribution is minor.

At larger $U$, as illustrated in Fig.~\ref{fig:eps}(b)--(c), the NRG results remain close to the linear expansion even for $\Gamma$ as large as $\Delta$. This behavior can be understood directly from the structure of the variational state. Away from a threshold, an $n$-QP component couples to the $(n-1)$-QP sector and first contributes to the $W=0$ energy at order $\Gamma^n$. Increasing the interaction suppresses the weights of the 1-QP states $\ket{\psi_2}, \ket{\psi_3}$ and consequently those of all components containing more QPs. Hence, higher-order contributions in $\Gamma$ become progressively less important as $U$ increases.

\begin{figure}[!htbp]
    \centering
    \includegraphics{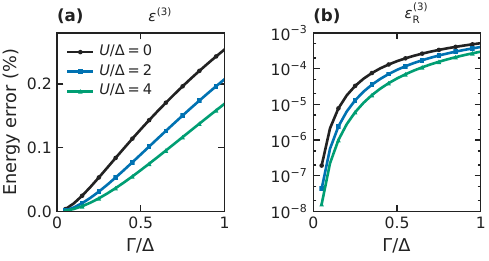}
    \caption{Accuracy of the level-3 approximation within the two-quasiparticle
    variational space at $\phi=0$, $D/\Delta=100$, and $W=0$.
    (a) Relative total-energy error
    $100[\epsilon^{(3)}-\epsilon_{\mathrm{2QP}}]/|\epsilon_{\mathrm{2QP}}|$
    of the scalar kernel closure.
    (b) Relative error
    $100[\epsilon_{\mathrm R}^{(3)}-\epsilon_{\mathrm{2QP}}]/|\epsilon_{\mathrm{2QP}}|$
    of the Rayleigh quotient evaluated with the normalized level-3 wavefunction
    constructed at $\epsilon^{(3)}$.
    Here $\epsilon_{\mathrm{2QP}}$ is the lowest energy of the full projected
    2-QP Hamiltonian. Both errors are expressed in percentages, with a linear
    vertical scale in (a) and a logarithmic vertical scale in (b).}
    \label{fig:kernel_accuracy_a}
\end{figure}

To test the level-3 approximation, we also solve the coupled integral equations \eqref{eq:alpha_final} and \eqref{eq:beta_final} numerically without approximating the kernel.
At $\phi=0$, for $U/\Delta=0,2,4$ and $0.05\leq\Gamma/\Delta\leq1$, the kernel-truncation relative error stays below $0.255\%$ [Fig.~\ref{fig:kernel_accuracy_a}(a)], increasing with $\Gamma$ and decreasing with $U$.
The kernel truncation is therefore not the main source of the differences from NRG in Fig.~\ref{fig:eps}. These come mainly from the omitted 3-QP and higher sectors. Appendix~\ref{truncation} demonstrates energy convergence at a representative parameter set and summarizes additional cutoff tests for the empty-dot probability and linear spin response.

A simple refinement is to evaluate the Rayleigh quotient of the level-3 wavefunction $\ket{\Psi^{(3)}}$, with all amplitudes constructed at the scalar root $\epsilon^{(3)}$:
\begin{equation}
    \epsilon_{\mathrm R}^{(3)} =
    \frac{\langle\Psi^{(3)}|H-E_{\mathrm{BCS}}^{(0)}|\Psi^{(3)}\rangle}
    {\langle\Psi^{(3)}|\Psi^{(3)}\rangle}.
    \label{eq:rayleigh_energy}
\end{equation}
Since this state belongs to the 2-QP variational space, the variational principle guarantees $\epsilon_{\mathrm R}^{(3)}\geq\epsilon_{\mathrm{2QP}}$, providing an upper bound to the lowest energy in the same symmetry sector.

The improved accuracy follows from the stationarity of the energy functional at the exact projected eigenstate: wavefunction errors affect the Rayleigh quotient only quadratically. For the level-3 approximation, this reduces the weak-coupling energy error from $O(\Delta g^3)$ to $O(\Delta g^5)$, where $g=\Gamma/\Delta$, away from thresholds. Over the parameter range shown in Fig.~\ref{fig:kernel_accuracy_a}(b), the relative error remains below $5.3\times10^{-4}\%$. 
Since the wavefunction obtained from the level-3 closure is more accurate than its scalar energy root $\epsilon^{(3)}$, re-evaluating the energy through the Rayleigh quotient makes fuller use of the information contained in the approximate state. In particular, the level-3 wavefunction incorporates both the diagonal 2-QP feedback and the leading mixed-kernel correction, which couples the $\alpha$ and $\beta$ sectors through 2-QP intermediate states.

\subsection{Wavefunction benchmarks and the local-moment fraction}

\begin{figure}[!htbp]
    \centering
    \includegraphics{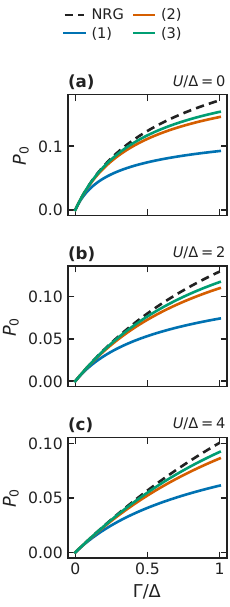}%
    \includegraphics{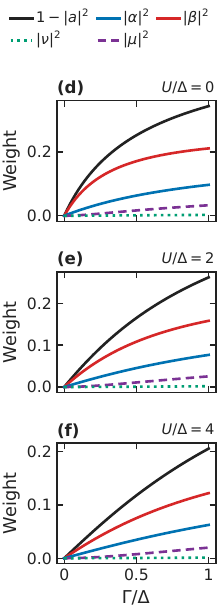}
    \caption{Wavefunction benchmarks at $\phi=0$. (a)--(c) Empty-dot probability $P_0$ versus $\Gamma$ for $U/\Delta=0,2,4$, respectively: NRG results and estimates from the normalized variational wavefunctions at approximation levels 1--3. (d)--(f) Weight outside the 0-QP sector, $1-|a|^2$, and individual 1-QP and 2-QP component weights, calculated from the normalized level-3 wavefunction.}
    \label{fig:PW}
\end{figure}

We further benchmark the accuracy of different approximation levels by computing the impurity occupation probabilities $P_n$ and comparing them with NRG results. Here, $n=0,1,2$ corresponds to an empty, singly occupied, or doubly occupied impurity level. Since $P_2 = P_0$ in the particle-hole-symmetric case, and $P_0+P_1+P_2=1$, it is sufficient to benchmark only one of the three. In Fig.~\ref{fig:PW}(a)--(c), we compare the empty-dot probability, $P_0 = (|\alpha|^2 + |\beta|^2)/2$. Including 2-QP components improves the 1-QP estimate, with level 3 closest to NRG among the three approximations shown; the remaining deviations grow toward $\Gamma\sim\Delta$. At $\Gamma=\Delta$, the level-3 result underestimates $P_0$ by $8.3$--$10.5\%$ over the displayed interaction values. Increasing $Q$ systematically reduces this discrepancy, as quantified in Appendix~\ref{truncation} and Fig.~{S2} of the Supplemental Material.

We also plot the weights of the wavefunction components in Fig.~\ref{fig:PW}(d)--(f). The 2-QP weights are small, but this alone does not establish the accuracy of the Ansatz: for some observables, intersector coherences can contribute at lower order than the sector probabilities. Accuracy must therefore be assessed separately for each observable.

\begin{figure}[!htbp]
    \centering
    \includegraphics{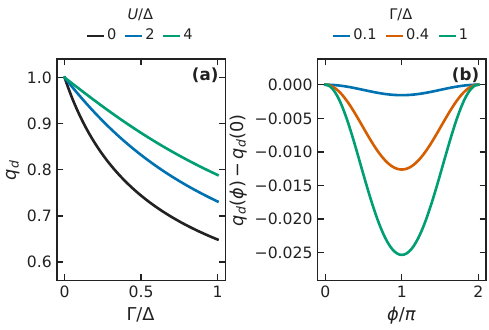}
    \caption{Local-moment fraction calculated from the normalized level-3 variational wavefunction at $W=0$. (a) $q_d=2\expval{S_d^z}$ versus $\Gamma$ at $\phi=0$ for several $U$ values. (b) Phase-dependent part $q_d(\phi)-q_d(0)$ versus $\phi$ at $U=2\Delta$ for several $\Gamma$ values.}
    \label{fig:szd}
\end{figure}

The local-moment fraction (i.e., the fraction of the total spin localized on the impurity) is
\begin{equation}
\begin{split}
q_d = 2\langle S^z_d \rangle &= |a|^2 + |\nu|^2 - \frac{1}{3} |\mu|^2 \\
&=1-P_0-P_2-\frac{4}{3}|\mu|^2 = P_1 - \frac{4}{3}|\mu|^2.
\end{split}
\label{eq:impurity_spin}
\end{equation}
The second line separates the loss of impurity spin through charge fluctuations from the additional compensation by the triplet pair. 
The negative bath-triplet contribution follows directly from the Clebsch--Gordan coefficients in Eq.~\eqref{eq:ansatz5}. This suppression of the on-site local moment (negative term) is a signature of partial Kondo screening in this system; the renormalization of the impurity spin (and the $g$ factor) has been discussed at length in Refs.~\cite{Moca2021Kondo,Pavesic2023Knight}. The rest of the spin resides in the leads.
Figure~\ref{fig:szd}(a) shows the impurity local-moment fraction $q_d$ as a function of $\Gamma$, while Fig.~\ref{fig:szd}(b) reveals a relatively weak dependence on $\phi$.

\section{Conventional and spin-dependent Josephson couplings}

In the odd-parity sector, the continuum is necessary to describe the Josephson current through the dot~\cite{Meng2009,Pokorny2023}; the first contribution occurs at the 2-QP level.

\subsection{Conventional Josephson energy}

\begin{figure}[!htbp]
    \centering
    \includegraphics{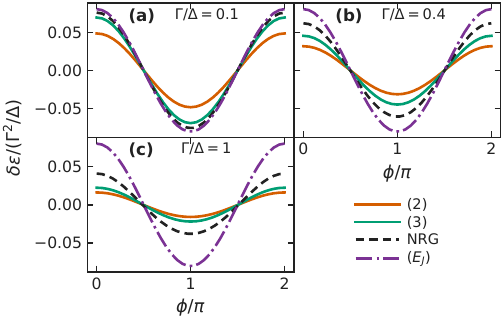}
    \caption{Shifted and rescaled doublet energy $\epsilon$ versus superconducting phase difference $\phi$ at $U = 2 \Delta$ for (a)--(c) $\Gamma/\Delta=0.1,0.4,1$. Solid: level-2 and level-3 approximations; dash-dotted: the leading low-$\Gamma$ result of Eqs.~\eqref{eq:EJ}--\eqref{eq:J_integral}, evaluated with the finite-band coefficient $\mathcal J_D(U/\Delta)$ defined in Eq.~\textup{(S20)}; dashed: NRG. Each curve is shifted by its phase average $\bar\epsilon=(2\pi)^{-1}\int_0^{2\pi}\epsilon(\phi)\,\dd\phi$, evaluated by trapezoidal integration with periodic closure on the supplied grids for the numerical curves, and rescaled by $\Gamma^2/\Delta$ ($\delta\epsilon=\epsilon(\phi)-\bar\epsilon$).}
    \label{fig:phi}
\end{figure}

In Fig.~\ref{fig:phi}, we show the doublet energy as a function of the phase difference $\phi$. The 1-QP result is omitted because it does not depend on $\phi$. By contrast, the 2-QP Ansatz correctly reproduces the leading $\cos{\phi}$ dependence and the negative first sine harmonic of the Josephson current $I=(2e/\hbar)\partial_\phi\epsilon$ ($\pi$-junction behavior)~\cite{vandam2006,rozhkov1999,clerk2000,Vecino2003,choi2004,siano2004,karrasch2008,Delagrange2015}. By expanding Eq.~\eqref{eq:2QP_app2} to second order in $\Gamma$, we recover the standard form~\cite{rozhkov2001,novotny2005} of the phase-dependent part of the energy
\begin{equation}
    \epsilon^{(\Gamma^2)} = E^{(\Gamma^2)}_J \cos\phi \>
    \label{eq:EJ}
\end{equation}
with the Josephson energy
\begin{equation}
    E_J^{(\Gamma^2)} = \frac{\Gamma^2}{\Delta} \mathcal{J}(u) \>,
    \label{eq:EJ2}
\end{equation}
where $u=U/\Delta$, and the dimensionless function is
\begin{equation}
    \begin{split}
        \mathcal{J}(u) &= \frac{1}{\pi^2} \int_1^\infty \frac{\dd x}{\sqrt{x^2-1}} \int_1^\infty \frac{\dd y}{\sqrt{y^2-1}} \\
        &\times \frac{1}{x+y}\bigg[\frac{1}{(x+\frac{u}{2})^2}+\frac{1}{(y+\frac{u}{2})^2}\\
        &\qquad\qquad+\frac{1}{(x+\frac{u}{2})(y+\frac{u}{2})}\bigg] \>.
    \end{split}
    \label{eq:J_integral}
\end{equation}

Equation~\eqref{eq:J_integral} is the particle-hole-symmetric specialization of the fourth-order perturbative result in Ref.~\cite{novotny2005}. Here it follows directly from the QP-sector expansion. All three terms describe virtual $0\to1\to2\to1\to0$-QP excursions at fourth order in the dot--lead hopping. The first two terms describe virtual transitions from a 1-QP configuration to a 2-QP intermediate and back to the same 1-QP configuration, and are related by $x\leftrightarrow y$. The last term describes transitions through a 2-QP intermediate that connect an $\alpha$ 1-QP configuration to a $\beta$ 1-QP configuration, or vice versa. The outer intermediate states have a charge-excited impurity and cost $\Delta(x+u/2)$ or $\Delta(y+u/2)$, while the middle intermediate state has a singly occupied impurity and costs $\Delta(x+y)$. Both singlet and triplet 2-QP configurations contribute.

The double integral in Eq.~\eqref{eq:J_integral} can be expressed in terms of elementary functions. We write $b=u/2$ to simplify notation.
For $b>-1$, we introduce
\begin{equation}
    F(b)=\begin{cases}
        \displaystyle\frac{\arccos b}{\sqrt{1-b^2}},&-1<b<1,\\[6pt]
        1,&b=1,\\[3pt]
        \displaystyle\frac{\operatorname{arcosh}b}{\sqrt{b^2-1}},&b>1.
    \end{cases}
    \label{eq:F_J}
\end{equation}
The result is
\begin{equation}
        \mathcal J(u)=\frac{1+3b}{4b(1+b)^2}+\frac{(3b^2-1)F(b)^2-2bF(b)}{\pi^2 b(1-b^2)} \>.
    \label{eq:J_closed}
\end{equation}
The apparent singularities at $b=0,1$ (i.e., $u=0,2$) are removable, with continuous values
\begin{equation}
    \mathcal{J}(0) = \frac{1}{4},\qquad
    \mathcal{J}(2) = \frac{1}{4}-\frac{5}{3\pi^2}.
\end{equation}
For large $U$,
\begin{equation}
    \mathcal{J}^{\mathrm{(large}\text{-}U)} \sim \frac{3}{u^2}, \quad \text{ for } 1\ll u\ll D/\Delta.
\end{equation}
In the noninteracting limit, the integral agrees with the weak-coupling odd-parity continuum energy
obtained in Ref.~\cite{kurilovich2021} [Eq.~(13)]. The large-$U$ asymptote agrees with the quantum-spin cotunneling result of Ref.~\cite{kirsanskas2015}, whose Eq.~(B35) yields $\mathcal{J}(u)\sim3/u^2$ at particle-hole symmetry and
with symmetric coupling. For $U/\Delta \ll 1$, the expansion reads
\begin{equation}
     E_J^{(\Gamma^2)} \approx
     \frac{\Gamma^2}{4\Delta}
     \left[
     1-\left(1-\frac{2}{\pi^2}\right)\frac{U}{\Delta}
     +\left(\frac54-\frac7{3\pi}\right)\left(\frac{U}{\Delta}\right)^2
     \right].
\end{equation}
The linear-$U$ correction agrees with Eq.~(B16) of Ref.~\cite{kurilovich2021}.

\begin{figure}[!htbp]
    \centering
    \includegraphics{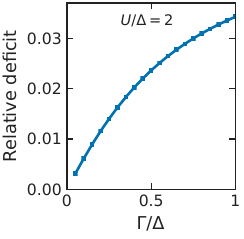}
    \caption{Kernel-truncation error in $E_J$: relative deficit $[E_{J,\mathrm{2QP}}-E_J^{(3)}]/E_{J,\mathrm{2QP}}$ of the first cosine harmonic from Eq.~\eqref{eq:2QP_app2}, at $U/\Delta=2$. Here $E_{J,\mathrm{2QP}}$ is the first cosine harmonic of $\epsilon_{\mathrm{2QP}}$, the lowest eigenvalue in the full 2-QP variational space.}
    \label{fig:kernel_accuracy_b}
\end{figure}

For small values of $\Gamma/\Delta$ [Fig.~\ref{fig:phi}(a)], the level-3 result and the second-order Josephson result $\epsilon^{(\Gamma^2)}$ approach the NRG benchmark for $E_J$, whereas the level-2 approximation omits part of the leading Josephson coefficient.
At order $\Gamma^2$, the additional term in $\epsilon^{(3)}$ supplies the mixed contribution in Eq.~\eqref{eq:J_integral}.
For large $\Gamma/\Delta$, $\epsilon^{(2,3)}$ underestimate $E_J$, while $\epsilon^{(\Gamma^2)}$ overestimates $E_J$. Figure~\ref{fig:kernel_accuracy_b} shows that the kernel-truncation error in $E_J$ is only a few percent, much smaller than the deviation from NRG at strong hybridization, which therefore stems mainly from the omitted 3-QP and higher configurations.
The leading kernel-truncation error, of order $\Gamma^3/\Delta^2$, is derived in Sec.~{IV\,B} of the Supplemental Material.

\subsection{Spin-dependent Josephson energy}

In the absence of the direct hopping term, the lowest odd-parity state is twofold degenerate, with spin-rotation partners $\ket{\Psi_{\up}}, \ket{\Psi_{\dn}}$. These states are constructed from the components in Eqs.~\eqref{eq:ansatz}--\eqref{eq:ansatz5}. The variational parameters for the spin-up case are given in Sec.~\ref{sec:ph-symmetric}. For the spin-down state defined by $\ket{\Psi_\dn}=S^-_{\mathrm{tot}} \ket{\Psi_\up}$, the 1-QP amplitudes are unchanged.
If the 2-QP operator polynomials are written by interchanging their spin labels, their coefficients obey $\mu_{\up} (\omega, \omega') = -\mu_{\dn} (\omega, \omega') = \mu(\omega, \omega')$ and $\nu_{\up} (\omega, \omega') = -\nu_{\dn} (\omega, \omega') = \nu(\omega, \omega')$.
In Eq.~\eqref{eq:H_T}, we chose the quantization axis along the preferred SOC axis. As a consequence, $H_T$ is diagonal in the $\{ \ket{\Psi_{\up}}, \ket{\Psi_{\dn}} \}$ basis, and the first-order corrections to the eigenenergies are simply $\delta \epsilon_{\sigma} = \expval{H_T}{\Psi_{\sigma}}$. Retaining matrix elements within the 1-QP sector and between the 0- and 2-QP sectors yields
\begin{equation}
    \delta \epsilon_{\sigma} \simeq - \sigma W_S \bigg[  \operatorname{Im}(A_s) A_a + B_s \operatorname{Im}(B_a)  + 2 a\operatorname{Im} M \bigg].
    \label{eq:deltaepsUP}
 \end{equation}
The first two terms, which come from the 1-QP components, are expressed in terms of
\begin{equation}
    \begin{split}
        A_s &= \rho \int \dd \omega \alpha^*(\omega) s(\omega), \quad
        A_a = \rho \int \dd \omega \alpha^*(\omega) a(\omega), \\ 
        B_s &= \rho \int \dd \omega \beta^*(\omega) s(\omega), \quad
        B_a = \rho \int \dd \omega \beta^*(\omega) a(\omega) \>.         
    \end{split}
\end{equation}
Here, we use the following properties of the 1-QP amplitudes: $\alpha(\omega)=-\alpha^*(-\omega)$ and $\beta(\omega)=\beta^*(-\omega)$. The last term comes from the coupling between the 0- and 2-QP components, where we introduce
\begin{equation}
    M =  \frac{\rho^2}{\sqrt{6}} \int \dd \omega \dd \omega' \mu^* (\omega, \omega') [ s (\omega) s (\omega') + a(\omega) a(\omega') ] \>.
\end{equation}
The 1-QP amplitudes $\alpha,\beta$ are of first order in $V$, whereas the 2-QP amplitudes $\mu,\nu$ are of second order. Expressed in Bogoliubov operators, $H_T$ contains both QP-number-conserving terms and terms that create or annihilate a pair. The former give the 1-QP contributions, while the latter produce the interference term $2a\operatorname{Im} M$. The triplet-sector weight is quadratic in $\mu$, whereas its interference with the 0-QP component is linear in $\mu$. The dashed $a$--$\mu$ link in Fig.~\ref{fig:setup}(c) represents this interference contribution. The 1-QP and 0--2-QP contributions therefore enter at the same order, $\mathcal{O}[(\rho W)\Gamma]$.
For the spin-up Ansatz, $\bra{\psi_5}H_T\ket{\psi_1}=iaW_SM$; including its Hermitian conjugate gives the interference term $-2aW_S\operatorname{Im}M$.

We therefore retain the $11$ and $02$ contributions in Eq.~\eqref{eq:deltaepsUP} and omit the $22$ contribution, in which both the bra and the ket contain two QPs. The latter starts at order $(\rho W_S)\Gamma^2/\Delta$ and therefore does not affect the leading-order result below; its role at finite $\Gamma$ is discussed in Sec.~\ref{sec:previous_works}.

Expanding the retained expression to leading order in $\Gamma$ and first order in $W$, with $\rho W\ll1$ and $D\gg U,\Delta$, we reproduce Eq.~\eqref{eq:phenom}, with
\begin{equation}
    \delta \epsilon_{\sigma} = \sigma E^{(\rho W \Gamma)}_S \sin{\phi} \>.
\end{equation}
Here $\delta \epsilon_{\sigma}$ denotes the part of the energy that depends on both $\phi$ and $\sigma$, and we introduced the quantity
\begin{equation}
    E^{(\rho W \Gamma)}_S = 2\rho W_S\Gamma\mathcal S(u),
\end{equation}
where the dimensionless integral for $u=U/\Delta\ge0$ is
\begin{equation}
\label{Su}
    \begin{split}
        \mathcal{S}(u) 
        &= \frac{1}{\pi} \int_{1}^\infty \int_{1}^\infty \frac{\dd x}{\sqrt{x^2-1}} \frac{\dd y}{\sqrt{y^2-1}} \\ 
        & \times \frac{(x+u/2) + (y+u/2) + (x+y) }{(x+u/2)(y+u/2)(x+y) } 
        = \frac{\pi}{2+u}.
    \end{split}
\end{equation}
In microscopic parameters, the result is
\begin{equation}
    \label{eq:ES_finite_gap}
    E_S^{(\rho W \Gamma)} = \pi\rho W_S\Gamma\,\frac{\Delta}{\Delta+U/2}.
\end{equation}
The denominator $\Delta+U/2$ is the threshold for a virtual charge excitation accompanied by a lead QP. The factor $\Delta/(\Delta+U/2)$ therefore gives the suppression relative to the superconducting-atomic-limit ($U,\Gamma\ll\Delta\ll D$) result 
\begin{equation}
\label{ES}
    E_S^\mathrm{(atomic)} = \pi \rho W_S \Gamma.
\end{equation}

\subsection{Background Josephson energy}
\label{sec:background_josephson}

The conventional Josephson coefficient calculated at $W=0$ describes
the active path; we denote it $E_{J,\mathrm{act}}$. It must be distinguished from
the net coefficient. At second order in direct interlead
hopping, virtual 2-QP excitations generate an additional
spin-independent contribution. In the wide-band, weak-tunneling limit,
the leading coefficient is
\begin{equation}
 E_{J,\mathrm{bg}}=-\pi^2(\rho W)^2\Delta,
 \label{eq:EJ_background}
\end{equation}
which is the Ambegaokar--Baratoff result~\cite{ambegoakar1963};
a derivation in our notation is given in Sec.~{II} of the Supplemental Material. The spin rotation cancels for
singlet-pair transfer, so this term depends on $W^2=W_N^2+W_S^2$ and
is independent of $\chi$ at fixed $W$. Its negative sign corresponds
to a conventional $0$-junction contribution, opposing the positive
odd-parity active-path coefficient. To leading order in the joint
weak-coupling expansion, the scalar coefficient entering
Eq.~\eqref{eq:phenom} is therefore
\begin{equation}
 \begin{split}
 E_{J,\mathrm{net}}&\simeq E_{J,\mathrm{act}}+E_{J,\mathrm{bg}}\\
 &\simeq\frac{\Gamma^2}{\Delta}\mathcal J(u)
       -\pi^2(\rho W)^2\Delta.
 \end{split}
 \label{eq:EJ_net}
\end{equation}
No mixed term of order $\Gamma\rho W_N$ appears: at particle-hole
symmetry, $\eta$ maps $W_N\to-W_N$, so the energy is even in $W_N$. Neglecting the background relative to the active Josephson response
requires
\begin{equation}
 r_{\mathrm{bg}}\equiv\frac{|E_{J,\mathrm{bg}}|}{E_{J,\mathrm{act}}}
 \simeq\frac{\pi^2(\rho W)^2}
 {(\Gamma/\Delta)^2\mathcal J(u)}\ll1,
 \label{eq:background_criterion}
\end{equation}
in addition to $\rho W\ll1$. With our convention $\rho=1/(2D)$,
\begin{equation}
 r_{\mathrm{bg}}\simeq\frac{\pi^2}{4\mathcal J(u)}
 \left(\frac{W}{\Gamma}\right)^2
 \left(\frac{\Delta}{D}\right)^2.
 \label{eq:background_bandwidth}
\end{equation}
For the fitted spin-dependent coupling $|E_S|/h=305$\,MHz of Ref.~\cite{pitavidal2023} and our reference parameters, Sec.~{III} of the Supplemental Material estimates $\rho W\simeq0.012$ and $r_{\mathrm{bg}}\simeq0.11$, well within the weak-background-tunneling regime.

\subsection{Intersector coherence and the spin response}

\begin{figure}[!htbp]
\centering
\includegraphics{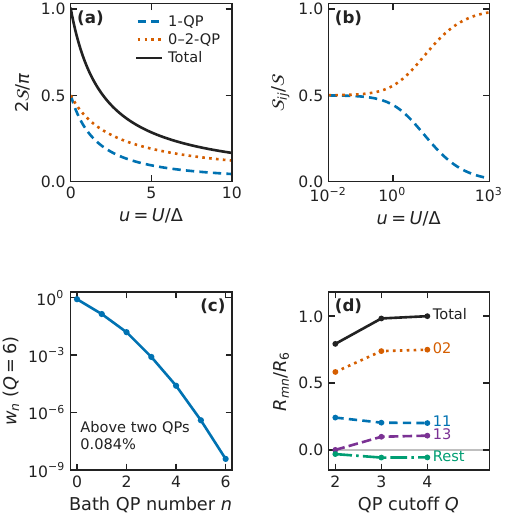}
\caption{QP-sector probabilities and coherence contributions to the spin response.
(a)--(b) Weak-hybridization, wide-band limit: (a) leading spin-coupling coefficients $2\mathcal S_{11}/\pi$, $2\mathcal S_{02}/\pi$, and their sum; (b) their fractions of the total coupling.
(c) Sector probabilities $w_n$ in the numerical $Q=6$ state.
(d) Signed response blocks $R_{mn}^{(Q)}$ and their total $R_Q$ for $Q=2,3,4$, normalized by $R_6$. Off-diagonal blocks include both Hermitian conjugates; ``Rest'' sums all blocks other than $11$, $02$, and $13$.
Panels (c)--(d) use $U/\Delta=2$, $\Gamma/\Delta=0.4$, $\phi=3\pi/5$, $D/\Delta=100$, and $W=0$, with 32 signed bath levels per lead.
}
\label{fig:sector_contributions}
\end{figure}

We can decompose $\mathcal S$ into two physical contributions, $\mathcal{S} = \mathcal S_{11}+\mathcal S_{02}$. In the kernel of Eq.~\eqref{Su}, the numerator $x+y$ gives the 1-QP term, while $(x+u/2)+(y+u/2)$ gives 0--2-QP interference. Recalling $b=U/(2\Delta)$ and $F(b)$ from Eq.~\eqref{eq:F_J}, the first contribution factorizes and has the closed form
\begin{align}
\mathcal S_{11}(u)&=\frac1\pi\left[\int_1^\infty
\frac{\dd x}{\sqrt{x^2-1}(x+b)}\right]^2
=\frac{F(b)^2}{\pi},
\end{align}
while the second term can be expressed as
\begin{equation}
    \mathcal S_{02}(u)=\frac{\pi}{2+u}-\mathcal S_{11}(u).
\label{eq:S02}
\end{equation}
At $U=0$, the two kernels coincide, so $\mathcal S_{11}(0)=\mathcal S_{02}(0)=\pi/4$. Thus 0--2-QP coherence supplies exactly half of $E_S$ at leading order. For $U>0$, the interference kernel exceeds the 1-QP kernel by the factor $1+u/(x+y)$, so its share is greater than one half; at $U=2\Delta$ it is $1-4/\pi^2\simeq0.595$. In the large-interaction regime $\Delta\ll U\ll D$,
\begin{equation}
\mathcal S_{11}(u)\sim\frac{4\ln^2u}{\pi u^2},
\qquad \mathcal S_{02}(u)\sim\frac{\pi}{u}.
\label{eq:S_sector_limits}
\end{equation}
The 1-QP term contains two charge-excitation denominators. The interference term contains only one, together with a neutral 2-QP excitation energy that does not include $U$. Increasing $U$ suppresses both contributions, the 1-QP term more strongly. The interference contribution therefore supplies the leading $1/U$ behavior, as illustrated in Fig.~\ref{fig:sector_contributions}(a)--(b).

At finite hybridization, the same mechanism extends to coherences involving higher-QP sectors. For the normalized numerical spin-up state at cutoff $Q$, we define the dimensionless response blocks
\begin{equation}
 R_{mn}^{(Q)}=\frac{2-\delta_{mn}}{\Delta}\operatorname{Re}
 \left\langle\Psi_{Q,m}\left|
 \frac{\partial H}{\partial(\rho W_S)}
 \right|\Psi_{Q,n}\right\rangle_{W=0},
 \label{eq:spin_response_blocks}
\end{equation}
where $m\leq n$ and $\ket{\Psi_{Q,n}}$ denotes the $n$-QP component of the same normalized state. Their sum, $R_Q=\sum_{m\leq n}R_{mn}^{(Q)}$, determines the signed spin splitting to linear order in $W_S$ through $\epsilon_\up-\epsilon_\dn=2\rho W_S\Delta R_Q$.

For the parameters of Fig.~\ref{fig:sector_contributions}(c)--(d), the $Q=6$ state has only
$0.084\%$ of its probability above two QPs, yet the full $Q=2$ response underestimates $R_6$ by
$20.8\%$. At $Q=4$, the $1$--$3$ coherence supplies $10.6\%$ of the total response, and the
$0$--$2$ contribution is enhanced relative to its $Q=2$ value. Higher-QP configurations therefore
contribute both through new coherences and through the dressing of amplitudes already present in
the two-QP state. Further details on convergence with respect to $Q$ are given in Appendix~\ref{truncation} and Sec.~{VIII\,B} of the Supplemental Material.

\subsection{Benchmarks and relation to previous results}
\label{sec:previous_works}

\begin{figure}[!htbp]
    \centering
    \includegraphics{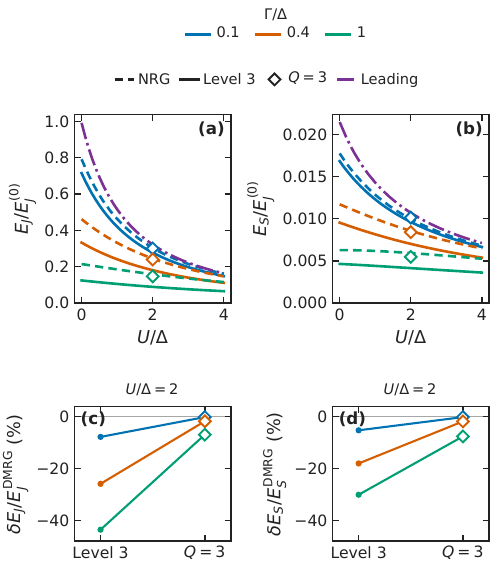}
    \caption{Josephson couplings and numerical benchmarks.
    (a)--(b) Couplings versus $U$, in units of $E_J^{(0)}=\Gamma^2/(4\Delta)$:
    (a) coefficient $E_J=[\epsilon(0)-\epsilon(\pi)]/2$ at $W=0$;
    (b) $E_S=|\epsilon_\up-\epsilon_\dn|/2$ at $\phi=\pi/2$, $W/\Gamma=0.4$, and $\chi=2\pi/3$.
    Dashed: NRG; solid: level-3 closure, using the $11+02$ response in (b); open diamonds: numerical $Q=3$ results at $U/\Delta=2$; purple dash-dotted: leading finite-band expressions from Eq.~\textup{(S20)}.
    (c)--(d) Relative deviations from DMRG at $U/\Delta=2$, with $\delta E=E-E^{\mathrm{DMRG}}$. Lines join the level-3 and $Q=3$ results for each hybridization. Colors denote $\Gamma/\Delta$ throughout; $D/\Delta=100$ in all panels.}
    \label{fig:EJES}
\end{figure}

Figure~\ref{fig:EJES}(a)--(b) compares $E_J$ and $E_S$ from the level-3 variational approximation
with NRG benchmarks for several values of $\Gamma$. DMRG calculations provide an independent
reference for the couplings in Fig.~\ref{fig:EJES}; the comparison is shown in
Fig.~{S3} of the Supplemental Material. On common $U$ grids, the NRG
values of $E_S$ differ from DMRG by $-2.5\%$ to $-0.9\%$ at $\Gamma/\Delta=0.1$, and by less than
$0.11\%$ at $\Gamma/\Delta=0.4$ and $1$.  The differences in $E_J$ are below $1.3\%$.

Both couplings decrease with increasing Coulomb repulsion. The leading weak-hybridization
expressions, Eqs.~\eqref{eq:EJ2} and \eqref{eq:ES_finite_gap}, overestimate their magnitudes,
increasingly so as $\Gamma/\Delta$ grows. The level-3 treatment captures this finite-hybridization
suppression but underestimates both couplings.

We attribute much of this deficit to incomplete dressing of the 2-QP intermediate states. The doublet energy is lowered by approximately $\Gamma G(U)/\pi$ [Eq.~\eqref{eq:expansion}], which contains the band logarithm $c_B$. The 2-QP configurations would acquire a comparable shift through their own charge fluctuations, which require 3-QP configurations. Within the 2-QP projection, the denominators $\epsilon-E(\omega)-E(\omega')$ are therefore too large in magnitude, and the finite-$\Gamma$ corrections acquire a spurious $\ln(D/\Delta)$ dependence (Sec.~{IV\,C} of the Supplemental Material).

Consistent with this picture, the numerical $Q=3$ results lie much closer to the DMRG reference [Fig.~\ref{fig:EJES}(c)--(d)]. The enlarged projected Hamiltonian is solved without a scalar-kernel closure, with finite background tunneling retained in the spin-splitting calculation. At $U/\Delta=2$ and $\Gamma/\Delta=0.4$, the deficits in $E_J$ and $E_S$ decrease from $25.9\%$ and $18.0\%$ for level 3 to $1.7\%$ and $1.8\%$ for $Q=3$. At $\Gamma=\Delta$, the remaining $Q=3$ deficits are $6.9\%$ and $7.6\%$, respectively, so higher-QP corrections are still appreciable. Numerical settings and bath-refinement checks are given in Appendix~\ref{truncation}.

The spurious bandwidth dependence provides a further diagnostic of the missing dressing. At $U=0$, where the model is quadratic, the exact couplings change by less than $1\%$ between $D/\Delta=10^2$ and $10^4$, whereas the level-3 $E_J$ at $\Gamma/\Delta=0.4$ decreases by $22\%$. A simple refinement within the two-QP space is to dress each pair denominator with the 1-QP self-energy of Eq.~\eqref{eq:1QP}, evaluated at the energy of the 2-QP configuration. This cancels the leading denominator mismatch, reduces the bandwidth dependence of $E_J$ and the retained $11+02$ response, and reduces their deviations from DMRG by factors of about $2$--$4$ and $1.5$--$3$, respectively (Sec.~{IV\,D} of the Supplemental Material). Dressing does not remove the bandwidth dependence of the omitted $22$ term, whose compensation requires 3-QP configurations. The finite-$\Gamma$ curves in Fig.~\ref{fig:EJES}(b), evaluated with the normalized level-3 wavefunction, therefore use the $11+02$ response as a partial-response approximation rather than the complete next-order spin response.

The interaction dependence differs between the two couplings, with stronger suppression of $E_{J,\mathrm{act}}$ compared to $E_S$. In the weak-hybridization limit, this distinction becomes explicit in the asymptotic laws $E_{J,\mathrm{act}}\propto U^{-2}$ and $E_S\propto U^{-1}$ for $\Delta\ll U\ll D$. Increasing the interaction therefore enhances the ratio $E_S/E_{J,\mathrm{act}}$ while reducing the absolute spin-dependent coupling. The background term is independent of $U$ at this order, so the total Josephson coefficient in Eq.~\eqref{eq:EJ_net} can change sign as the active contribution decreases. 

Our expression for $E_S$ has the same dependence on the interaction strength and the superconducting gap as the cotunneling result of Padurariu and Nazarov~\cite{padurariu2010}, evaluated in the limit where a high-energy dot orbital can be integrated out. For this contribution, our explicit evaluation yields twice the magnitude of the Pauli-operator coefficient in their Eq.~(23). Section~{V} of the Supplemental Material details the correspondence between the models, the conventions, and the normalization underlying this comparison. It traces the factor of two to the addition of the forward and reverse transfer amplitudes, as in their Eq.~(19).
Our normalization is supported independently by the atomic-limit agreement discussed below and by the NRG benchmarks in Fig.~\ref{fig:EJES}(b). At $\Gamma/\Delta=0.1$, the NRG values lie between $82\%$ and $94\%$ of Eq.~\eqref{eq:ES_finite_gap} for $0\leq U/\Delta\leq4$; the remaining difference is the finite-$\Gamma$ suppression discussed above.

In the superconducting atomic limit $U,\Gamma\ll\Delta\ll D$, Eq.~\eqref{eq:ES_finite_gap} reduces to Eq.~\eqref{ES}, which agrees with the weak-background-tunneling result of Bargerbos et al.~\cite{bargerbos2023}. Section~{VI} of the Supplemental Material establishes this agreement by accounting for the different spin bases and hybridization conventions. Equation~\eqref{eq:ES_finite_gap} thus extends the atomic-limit expression from that work to finite $U/\Delta$.

\section{Magnetic Josephson response}
\label{sec:magnetic_response}

The spin splitting calculated above does not specify how the spin is distributed between the dot and the leads. We calculate how this distribution changes when SOC is included and show that a magnetic field can turn this change into an additional Josephson current. We set $W_N=0$ and work to first order in $\rho W_S$. In the absence of a magnetic field, the odd-parity states change from $|\Psi_\sigma^{(0)}\rangle$ at $W=0$ to $|\Psi_\sigma\rangle$ at $W_S \neq 0$. Since $H_T$ conserves $S^z_{\mathrm{tot}}$, the label $\sigma=\pm1$ continues to specify $S^z_{\mathrm{tot}}=\sigma/2$.

\subsection{SOC-induced spin transfer between dot and leads}

The expectation value of the dot spin operator in each state defines the corresponding local-moment fraction,
\begin{equation}
 q_{d,\sigma}\equiv2\langle S_d^z\rangle_\sigma
 =2\langle\Psi_\sigma|S_d^z|\Psi_\sigma\rangle.
 \label{eq:mag_dot_spin}
\end{equation}
The factor of two gives $q_{d,\sigma}=\sigma$ when the entire spin
resides on the dot. At $W=0$, spin-rotation symmetry instead gives
$q_{d,\sigma}=\sigma q_d^{(0)}$, where $q_d^{(0)}$ is the local-moment fraction
from Eq.~\eqref{eq:impurity_spin}. We now show that SOC breaks
the relation $q_{d,\uparrow} = -q_{d,\downarrow}$.

Introducing a generator field (auxiliary local Zeeman term)
$H_\lambda=\lambda(n_\up-n_\dn)=2\lambda S_d^z$, we obtain from the Hellmann--Feynman theorem 
\begin{equation}
 q_{d,\sigma}
 =\left.\frac{\partial\epsilon_\sigma(\lambda)}{\partial\lambda}\right|_{\lambda=0}.
 \label{eq:mag_hellmann_feynman}
\end{equation}

The leading contribution to $E_S$ comes from processes containing two dot--lead hops
and one direct interlead hop. When the dot is temporarily empty or
doubly occupied, its spin is zero. Relative to the initial dot energy
$\sigma\lambda$, the cost of this charge excitation is therefore
$\Delta+U/2-\sigma\lambda$. A 2-QP intermediate has a singly occupied
dot. In the processes contributing at this order, $H_T$
creates or annihilates the QP pair without changing the dot spin.
The dot Zeeman energies then cancel in the excitation cost.
Thus, in the leading spin-dependent energy, adding the generator
field is equivalent to replacing $U$ by $U-2\sigma\lambda$:
\begin{equation}
 \epsilon_\sigma^{(\rho W \Gamma)}(\lambda)
 =\sigma E_S(U-2\sigma\lambda)\sin\phi.
 \label{eq:mag_probe_energy}
\end{equation}
The SOC-induced correction is the same for both states:
\begin{equation}
 \begin{aligned}
 q_{d,\sigma}&=\sigma q_d^{(0)}+\delta q_d+\mathcal O\left((\rho W_S)^2\right),\\
 \delta q_d&\simeq-2\frac{\partial E_S}{\partial U}\sin\phi
     =\frac{E_S\sin\phi}{\Delta+U/2}.
 \end{aligned}
 \label{eq:mag_spin_transfer}
\end{equation}
The last equality uses Eq.~\eqref{eq:ES_finite_gap}. This is the
leading weak-hybridization result.

For $W_S\sin\phi>0$, the dot spin in the $\sigma=+1$ state becomes
more positive, while that in the $\sigma=-1$ state becomes less
negative. In each state the total spin projection remains $\sigma/2$, so the leads acquire
the opposite change in their combined spin expectation value.

The SOC-induced corrections to the wavefunction require an extension of the
Ansatz by spin-quartet components ($S=3/2$, $S_z=1/2$).

\subsection{Common energy shift and Josephson current}

We now assume that a weak magnetic field is applied along the SOC axis. Let $b_d$ be the Zeeman energy on the dot and
$b_L=b_R=b_{\mathrm{lead}}$ the common Zeeman energy in the leads. With
$b_{\mathrm{rel}}=b_d-b_{\mathrm{lead}}$, the magnetic term is
\begin{equation}
 H_Z=b_dS_d^z+b_{\mathrm{lead}}S_{\mathrm{bath}}^z
    =b_{\mathrm{lead}}S_{\mathrm{tot}}^z+b_{\mathrm{rel}}S_d^z.
 \label{eq:mag_zeeman}
\end{equation}
The linear field-induced energy shift is
\begin{equation}
 \Delta_B\epsilon_\sigma
 =\frac{\sigma}{2}\left[b_{\mathrm{lead}}+b_{\mathrm{rel}}q_d^{(0)}\right]
  +\frac{b_{\mathrm{rel}}\delta q_d}{2}.
 \label{eq:mag_effective}
\end{equation}
The first term shifts the two levels in opposite directions and
therefore changes their splitting. It corresponds to the familiar
renormalization of the magnetic response when part of the spin
resides in the leads~\cite{Pavesic2023Knight}. The term containing
$\delta q_d$ instead shifts both levels equally. Because this shift
depends on the superconducting phase difference, the arithmetic
mean energy $\bar\epsilon_{\mathrm{spin}}=(\epsilon_\up+\epsilon_\dn)/2$ and the corresponding mean
current $\bar I=(I_\up+I_\dn)/2$ acquire corrections
\begin{equation}
 \delta\bar\epsilon_{\mathrm{spin}}=\frac{b_{\mathrm{rel}}E_S\sin\phi}{2(\Delta+U/2)}, \quad
 \delta\bar I=\frac{e\,b_{\mathrm{rel}}E_S\cos\phi}{\hbar(\Delta+U/2)}.
 \label{eq:mag_common_current}
\end{equation}
This contribution cancels from the qubit transition energy but survives in the mean Josephson current, revealing spin redistribution that the splitting alone does not probe. At zero phase bias, the current correction is nonzero for $b_{\mathrm{rel}}E_S\ne0$ and reverses sign with $b_{\mathrm{rel}}$.
Supercurrents at zero phase bias are the hallmark of the anomalous Josephson ($\varphi_0$) effect, produced by SOC combined with broken time-reversal symmetry~\cite{Buzdin2008,Reynoso2008,zazunov2009,Brunetti2013,Yokoyama2014} and observed in nanowire quantum dots~\cite{Szombati2016}. In Eq.~\eqref{eq:phenom}, each qubit state already carries the spin-dependent current $\sigma(2e/\hbar)E_S\cos\phi$, so a field that polarizes the qubit produces such a current through the state populations. The contribution in Eq.~\eqref{eq:mag_common_current} is different. It survives in the average over both states and requires $b_d\neq b_{\mathrm{lead}}$, because a uniform Zeeman term $b_{\mathrm{lead}}S^z_{\mathrm{tot}}$ shifts the two states by $\pm b_{\mathrm{lead}}/2$ without phase dependence. It therefore measures the SOC-induced spin transfer $\delta q_d$.
The common energy shift could also be probed using occupation-resolved microwave susceptibility
\cite{Fatemi2022Susceptibility}.

\section{Spatial structure in the superconducting leads}
\label{sec:real_space}

We now distinguish the localization of virtual QPs~\cite{pavesic2024} from the physical spin distribution~\cite{Moca2021Kondo,Pavesic2023Knight}, which also contains interference between QP sectors. We calculate both by solving the full 2-QP problem at $W=0$ without making any kernel approximations~\cite{BCSQPSolver2026}.

In the impurity model, the energy-dependent hybridization function fully describes the dot--lead coupling, and the wavefunctions we obtain are expressed through energy-dependent coefficients. To probe the real-space properties, we additionally specify the shape of the single-particle wavefunctions in the leads. We make this choice in a way that preserves the flat normal-state local density of states (LDOS) exactly, with constant hopping from the dot to the first site of each superconducting chain. This construction is an application of the orthogonal-polynomial continuum-to-chain mapping~\cite{chin2010,devega2015,nuesseler2020}, specialized here to superconducting leads. Since our aim is to capture qualitative trends, we choose this representation for convenience. Extending the framework to a specific device with energy-dependent hybridization should be straightforward.

\subsection{Chain with exactly flat local density of states}

For the normalized measure $\rho \, \dd \omega = \dd\omega/(2D)$, we introduce the orthonormal Legendre functions
\begin{equation}
    p_n(\omega) = \sqrt{2n+1}\,P_n(\omega/D), \qquad
    \int_{-D}^{D} \rho \, \dd \omega \, p_n p_m=\delta_{nm}.
    \label{eq:rs_legendre}
\end{equation}
Here $P_n$ is the ordinary Legendre polynomial and $n=0,1,\ldots$. Its three-term recurrence (Bonnet's formula) expresses multiplication by the normal-state energy $\omega$ as nearest-neighbor hopping:
\begin{equation}
    \omega p_n=t_n p_{n+1}+t_{n-1}p_{n-1},\qquad
    t_n=\frac{D(n+1)}{\sqrt{(2n+1)(2n+3)}},
    \label{eq:rs_hopping}
\end{equation}
with $t_{-1}=0$. The recurrence is equivalent to Lanczos tridiagonalization of the bath Hamiltonian, which acts as multiplication by energy. The corresponding electron-chain Hamiltonian is
\begin{equation}
        H_l= \sum_{n\sigma}t_n (c^\dag_{l,n+1,\sigma}c_{ln\sigma}+\mathrm{H.c.}) 
                -\Delta\sum_n(c^\dag_{ln\up}c^\dag_{ln\dn}+\mathrm{H.c.}).
    \label{eq:rs_chain}
\end{equation}
Here $H_l$ is the lead Hamiltonian $H_{\mathrm{BCS},l}$ expressed in the chain basis. The on-site singlet pairing remains uniform because the transformation is real and orthogonal, and the same transformation acts on the two time-reversed partners. The dot couples only to the first site of the chain, and the normal-state LDOS at this site is exactly $\rho_0(\omega)=1/(2D)$ for $|\omega|<D$. Replacing every $t_n$ by $D/2$ would instead give a semicircular LDOS.

Assigning a lattice spacing $a_0$ to this lead description gives $r_n = n a_0$. Since $t_n \to D/2$ as $n\to \infty$, the asymptotic dispersion of the chain is $\omega(k)=D\cos(ka_{0})$ and the Fermi velocity at half filling is $v_F = D a_{0} / \hbar$. Thus
\begin{equation}
    \xi_0 \equiv \frac{\hbar v_F}{\pi\Delta} = \frac{D}{\pi\Delta} a_{0},
    \label{eq:rs_length}
\end{equation}
where $\xi_0$ is the BCS coherence length~\cite{tinkham}. The parameters $\Gamma, D, \Delta$ fix the dimensionless couplings and the distances in units of $\xi_0$, but $\xi_0$ (or equivalently $a_{0}$ or $v_F$) has to be specified to discuss absolute lengths.

\subsection{Quasiparticle natural orbitals and localization}

\begin{figure}[!htbp]
 \centering
 \includegraphics{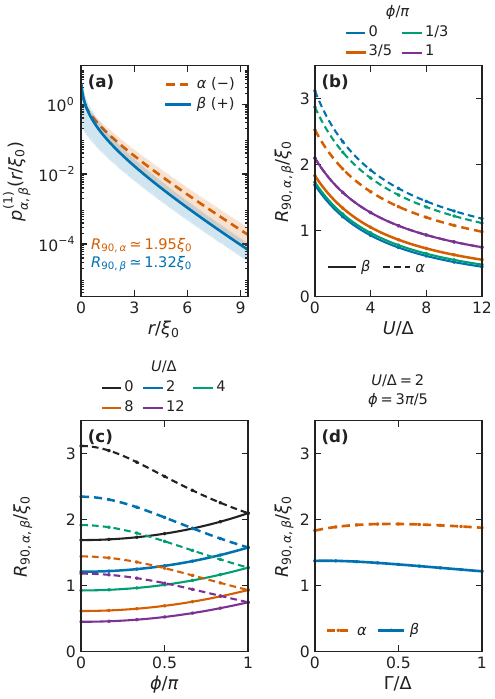}
 \caption{Spatial profiles and containment radii of the 1-QP components
 $\alpha$ ($-$ channel, dashed) and $\beta$ ($+$ channel, solid)
 in the full 2-QP variational state at $W=0$.
 (a) Unit-normalized densities $p_{\alpha,\beta}^{(1)}(r/\xi_0)$ at the reference point
 $U/\Delta=2$, $\Gamma/\Delta=0.4$, $\phi=3\pi/5$; component weights
 $|\alpha|^2=0.0520$ and $|\beta|^2=0.0839$.
 (b)--(d) $R_{90}$ versus $U/\Delta$, $\phi/\pi$, and $\Gamma/\Delta$,
 respectively; other parameters take their reference values unless indicated.
 In (c), $|\alpha(r;2\pi-\phi)|=|\beta(r;\phi)|$ [Eq.~\eqref{eq:phase_reflection}]: beyond
 $\phi=\pi$, each dashed curve continues as the mirror image of the solid curve with the same $U$.
 For this figure and Figs.~\ref{fig:real_space_two_qp} and~\ref{fig:real_space_profiles},
 distances are expressed in $r/\xi_0$ and adjacent sites form cells of width
 $h=2a_0/\xi_0=2\pi\Delta/D$; their summed weights are divided by $h$ to give densities.
 Faint lines show raw-site densities.
 $R_{90}$ encloses $90\%$ of each orbital's full probability
 and is expressed in units of $\xi_0$.
 }
 \label{fig:real_space_one_qp}
\end{figure}

In the chosen representation, the real-space amplitudes can be expressed as
\begin{equation}
    \begin{split}
        f(r_n) &= \int\rho\,\dd\omega\,p_n(\omega)f(\omega) \\
        F(r_n, r_m) &= \iint\rho^2\,\dd\omega\,\dd\omega'\, p_n(\omega)p_m(\omega')F(\omega,\omega'),
    \end{split}
 \label{eq:rs_coefficients}
\end{equation}
where $f \in \{\alpha, \beta\}$, $F \in \{\mu, \nu \}$, and all energy integrals run over $[-D,D]$. The transformation preserves the integrated component weights defined in Eq.~\eqref{eq:component_weights}. Figure~\ref{fig:real_space_one_qp}(a) shows the unit-normalized probability density of 1-QP components
\begin{equation}
    p_f^{(1)} (r_n) = \frac{|f (r_n)|^2}{|f|^2}
\end{equation}
in real space. We characterize the orbital widths by $R_{90,f}$, the radius enclosing $90\%$ of
the probability corresponding to $f$. The spatial extent of both channels is of the order of the
coherence length, but it varies with $U$, $\phi$, and $\Gamma$ [see also
Fig.~\ref{fig:real_space_one_qp}(b)--(d)]. Because the transformation in
Eq.~\eqref{eq:rs_coefficients} is real and independent of $\phi$, Eq.~\eqref{eq:phase_reflection}
also holds for $f(r_n)$ and $F(r_n,r_m)$. The $\alpha$ and $\beta$ profiles therefore coincide at
$\phi=\pi$ and are exchanged by reflection about this point, which explains the phase dependence in panel (c).

Both 2-QP components have one QP in the $+$ and one QP in the $-$ channel, with the corresponding coordinates $r_n$ and $r_m$. To characterize one QP without specifying the state of the other, we trace out all other degrees of freedom: the dot, the other QP, and the spin of the retained QP. The remaining orbital states are described by a channel-resolved 1-QP density matrix (see Eq.~\textup{(S58)} of the Supplemental Material for a derivation),
\begin{equation}
    \begin{aligned}
        (\varrho_+)_{nn'} &= \frac1{w_2} \sum_m (\nu_{nm} \nu_{n'm}^*+\mu_{nm}\mu_{n'm}^*), \\
        (\varrho_-)_{mm'}&= \frac1{w_2} \sum_n (\nu_{nm}\nu_{nm'}^*+\mu_{nm}\mu_{nm'}^*).
    \end{aligned}
\label{eq:rs_natural_orbitals}
\end{equation}

Here we used the abbreviation $F(r_n, r_m) = F_{nm}$ and normalized the density matrices by the total 2-QP weight $w_2 = |\mu|^2 + |\nu|^2$.
The diagonal elements $(\varrho_\pm)_{nn}$ give the probability for the QP to occupy site $n$, while the off-diagonal elements describe coherence between different orbitals. Both density matrices are Hermitian and positive semidefinite, with unit trace because the selected channel contains exactly one QP. The eigenvectors of $\varrho_{s}$ are the natural orbitals $\psi_{s,j}(r_n)$ of the normalized 2-QP component, and the corresponding eigenvalues give their occupation probabilities $\lambda_{s,j}$. An occupation close to one means that a single wavepacket captures most of that QP's orbital weight. Related natural-orbital approaches have been used to expose compact orbital structures in quantum-impurity states. Yang and Feiguin~\cite{yang2017} introduced projected natural orbitals to identify a dominant screening orbital in finite Kondo systems. Debertolis, Florens, and Snyman~\cite{debertolis2021,debertolis2022} used natural-occupation spectra to quantify the few-body character of correlated impurity ground states. Here the natural orbitals characterize the normalized virtual 2-QP component of a superconducting doublet.

\begin{figure}[!tb]
 \centering
 \includegraphics{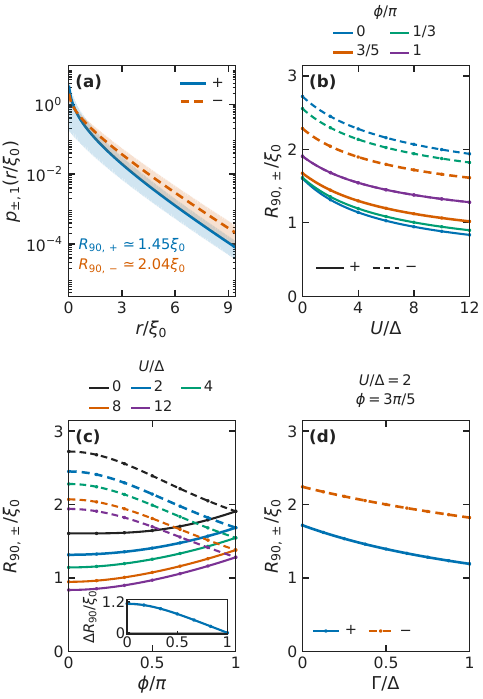}
 \caption{Leading 2-QP natural orbitals, with the panel arrangement, parameters,
 and conventions of Fig.~\ref{fig:real_space_one_qp}.
 (a) Densities $p_{\pm,1}(r/\xi_0)$ with
 occupations $\lambda_{+,1}=0.9767$ and $\lambda_{-,1}=0.9684$;
 the total 2-QP weight is $w_2=0.01081$.
 In (c), reflection about $\phi=\pi$ exchanges the $+$ and $-$ orbitals
 [Eq.~\eqref{eq:phase_reflection}], as in Fig.~\ref{fig:real_space_one_qp}(c).
 Inset in (c): $\Delta R_{90}=R_{90,-}-R_{90,+}$ for $U/\Delta=2$;
 the gray band spans the five displayed $U$ cuts.
 }
 \label{fig:real_space_two_qp}
 \label{fig:orbital_radii}
\end{figure}

Figure~\ref{fig:real_space_two_qp}(a) shows the unit-normalized probability densities
\begin{equation}
    p_{s, 1} (r_n) = \frac{|\psi_{s,1} (r_n)|^2}{|\psi_{s,1}|^2}
\end{equation}
of the leading orbital in each channel. Their high occupation probabilities $\lambda_{+,1}=0.98$
and $\lambda_{-,1}=0.97$ suggest that the pair is predominantly composed of one wavepacket per
channel. We again characterize the widths of the orbitals with the radii $R_{90,s}$, which are of the order $\xi_0$ as in the 1-QP case. Further details on the natural orbital analysis are given in Sec.~{VII\,C} of the Supplemental Material.

Figure~\ref{fig:orbital_radii}(b)--(d) shows the parameter dependence of the radii $R_{90,s}$ enclosing $90\%$ of the leading orbital probability. Panels (b) and (c) use $\Gamma/\Delta=0.4$, while panel (d) varies $\Gamma/\Delta$ at $U/\Delta=2$ and $\phi=3\pi/5$. At every sampled point, the occupations of both leading natural orbitals exceed $0.956$, so the
two-orbital description remains useful throughout these parameter sweeps. Increasing $U$ at fixed
phase contracts both orbitals [panel (b)]. By Eq.~\eqref{eq:phase_reflection}, $\varrho_\pm(2\pi-\phi)=\varrho_\mp^*(\phi)$, so the phase
dependence [panel (c)] has the same reflection symmetry as in the 1-QP case. The difference in spatial extent depends only weakly on $U$ over the displayed cuts and vanishes at $\phi=\pi$; see the inset to panel (c). At $\phi=\pi$ the spatial densities coincide, while the orbitals remain orthogonal through their channel indices.

\subsection{Physical spin density and dot--lead correlations}

To locate the physical spin, we evaluate
\begin{equation}
 q_l(n)=2\langle s_l^z(n)\rangle,\qquad
 C_l(n)=\langle\mathbf S_d\cdot\mathbf s_l(n)\rangle,
 \label{eq:rs_spin_definition}
\end{equation}
where $\mathbf s_l(n)=c^\dag_{ln}\boldsymbol\sigma c_{ln}/2$ and
$C_l$ is a dot--lead spin correlation function. 
The observables satisfy the sum rules
\begin{equation}
 q_d+\sum_{ln}q_l(n)=1,\qquad
 \sum_{ln}C_l(n)=-|\mu|^2.
 \label{eq:rs_sum_rules}
\end{equation}
The second identity follows because the bath triplet coupled to the dot
has $\langle \mathbf S_d\cdot\mathbf S_{\mathrm{bath}} \rangle = -1$, whereas the bath
singlet contributes zero.

\begin{figure}[!htbp]
 \centering
 \includegraphics{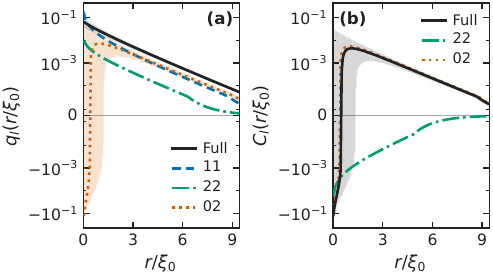}
 \caption{Zero--two-QP interference redistributes the physical spin without changing the integrated lead spin. Results are at the reference point of
 Fig.~\ref{fig:real_space_one_qp}.
 (a) Spin density $q_l(r/\xi_0)$ per lead, with its $11$, $22$, and $02$ contributions.
 (b) Dot--lead correlation $C_l(r/\xi_0)$, with its $22$ and $02$ parts.
 Here $11$ and $22$ denote contributions within the 1- and 2-QP sectors,
 while $02$ denotes combined vacuum--2-QP interference.
 Both ordinates are symmetric-logarithmic, linear for $|q_l|,|C_l|<10^{-4}/\pi$.
 }
 \label{fig:real_space_profiles}
\end{figure}

At the reference parameter set, $q_d\approx0.85$, and each lead carries $\sum_nq_l(n)\approx0.075$. The spin profile is positive, but the local
dot--lead correlation changes sign [Fig.~\ref{fig:real_space_profiles}(a) and (b)].
The normalized cumulative correlation initially exceeds its asymptotic value, reaching $3.09$ at $r\simeq0.44\xi_0$, before a positive tail compensates the short-distance negative correlation. The $02$ term contributes zero to the full spatial integral but dominates this redistribution.

\section{Conclusion}

We developed a continuum variational description of the odd-parity doublet in an interacting quantum-dot Josephson junction with spin-dependent background tunneling and assessed its accuracy against NRG and DMRG benchmarks. 
We compared our results with previous work and verified agreement in several limits.
Kernel truncation yields explicit equations for the energies and wavefunctions within the 2-QP projection and captures the leading conventional and spin-dependent Josephson couplings. 
The 0--2-QP interference contributes at the same leading order as the 1-QP sector despite the smaller 2-QP probability: it supplies exactly half of the spin-dependent coupling at $U=0$ and its dominant part for $\Delta\ll U\ll D$.
Higher-QP coherences extend this mechanism to finite-hybridization corrections, so convergence must be checked separately for each observable.
SOC also transfers spin between the dot and the leads. When the dot and lead Zeeman energies differ, this transfer produces a Josephson-current contribution that is common to both qubit states and persists at zero phase bias.
In real space, the virtual QPs extend over the coherence length $\xi_0$, and each channel of the 2-QP component is dominated by a single natural orbital; 0--2-QP interference redistributes the physical spin in the leads without changing its integrated value.

Our results provide microscopic constraints relevant to the design of superconducting circuits incorporating ASQs. In the weak-hybridization regime, $E_S$ is reduced from its superconducting-atomic-limit value by the factor $\Delta/(\Delta+U/2)$. This correction directly affects spin-dependent currents and circuit-mediated interactions. The Coulomb interaction suppresses the active-path conventional and spin-dependent Josephson contributions differently: for $\Delta\ll U\ll D$, they decay as $U^{-2}$ and $U^{-1}$, respectively. The growing magnitude of $E_S/E_{J,\mathrm{act}}$ thus reflects preferential suppression of the conventional coupling, not an increasing spin splitting. The gain in relative spin dependence comes at the cost of a smaller absolute spin-dependent coupling.

\begin{acknowledgments}
We thank Luka Pavešić, Don Rolih, Fabian Hassler, and \'Ad\'am B\'acsi for discussions and comments.
We acknowledge the support of the Slovenian Research and Innovation Agency (ARIS) under research program No. P1-0416.
\end{acknowledgments}

\section*{Data availability}

The numerical data underlying the figures and tables, calculation
parameters, and manuscript-specific analysis and figure-generation
scripts are openly available on Zenodo~\cite{ASQData2026}.
The \texttt{qdjj\_solver} software is available at
\url{https://github.com/rokzitko/qdjj_solver}, with a versioned
release archived on Zenodo~\cite{BCSQPSolver2026}.

\appendix

\section{Numerical methods}
\label{newappA}

\subsection{QP expansion}
\label{appAqp}

We implemented a highly efficient numerical impurity solver which diagonalizes
$H_Q=\mathcal{P}_Q H \mathcal{P}_Q$, where the projector $\mathcal{P}_Q$ retains all impurity
configurations and at most $Q$ bath QPs, defined relative to the isolated-lead
BCS vacuum. 
We use Gauss--Legendre quadrature in $t$, with $\omega=\Delta\sinh t$. Direct discretization of 
the five components with 32 nodes gives matrix dimension 2113 at $Q=2$.
Higher-$Q$ calculations enlarge the retained configuration space.
We obtain the lowest projected eigenstates by sparse iterative diagonalization. For a fixed
discretized Hamiltonian, these eigenenergies are variational upper bounds in the selected symmetry
sector.
At particle-hole symmetry ($\delta=0$) and $W=0$, we use this solver at $Q=2$ 
to test the scalar-kernel closure and by increasing $Q$ we test the contributions of omitted QPs. 
We can also include finite $W_N,W_S$ directly and retain both $\eta$ sectors when $\delta\neq0$ or $W_N\neq0$, 
extending beyond the five-component Ansatz from Sec.~\ref{sec:ph-symmetric}. We compute eigenvalues and eigenvectors,
dot-occupation probabilities, Josephson couplings, QP weights, and the QP-sector matrix elements
entering the spin response $R_{mn}^{(Q)}$.
The following Appendix~\ref{truncation} reports QP-cutoff and bath-refinement convergence properties.
The code is available as part of the \texttt{qdjj\_solver} package~\cite{BCSQPSolver2026}.

\subsection{DMRG}
\label{appAdmrg}

We obtain independent references using DMRG~\cite{White1992DMRG}, based on TeNPy
1.1.1~\cite{Hauschild2024TeNPy}, without a bath-QP-number cutoff. After exact
canonical transformations, we arrange the impurity and bath modes of a
star Hamiltonian along a matrix-product state (MPS). For spin splittings at finite $W$, 
we first diagonalize the quadratic Hamiltonian of the
coupled reservoirs, retaining the isolated-lead BCS energy reference
$E_{\mathrm{BCS}}^{(0)}$. We target the lowest odd-parity state with
$S_z=1/2$, and also $S_z=-1/2$ when calculating spin splittings. The common
bath quadrature permits direct comparisons with the QP expansion on identical
discretizations, which provides information about the QP-truncation effects 
once MPS convergence is established. We check bath resolution, MPS bond dimension, 
and sweep convergence separately for the energies, dot occupations, and Josephson
couplings.
This code is also available as part of the \texttt{qdjj\_solver} package~\cite{BCSQPSolver2026}.

\subsection{NRG}
\label{app:nrg}

We used the NRG Ljubljana code (version 2026.09)~\cite{NRGLjubljana2026}, with the discretization scheme from Ref.~\cite{ZitkoPruschke2009}.
At $\phi=0$ and $W=0$, the problem effectively has a single channel, allowing efficient NRG computations with discretization parameter $\Lambda = 2$ and $10\,000$ retained states. We used spin $\mathrm{SU}(2)$ symmetry and did not perform $z$-averaging. At $\phi \neq 0$ but $W=0$, the problem still has spin $\mathrm{SU}(2)$ symmetry, and we performed the calculations at $\Lambda = 4$ with $3000$ retained states. We averaged over $z\in \{ 0.75, 1.00 \}$, and the maximum absolute discrepancy between the results for the two $z$ values did not exceed $10^{-4} \Delta$.
For $\phi\neq0$, $W\neq0$, only spin $\mathrm{U}(1)$ symmetry remains; we used $\Lambda=8$, retained $3000$ states, and averaged over $z\in \{0.5, 1\}$. The spin-splitting calculations used $B_z=10^{-8}D=10^{-6}\Delta$.
This field slightly biases the extracted $E_S$ when the splitting is small. 
The absolute energies in Fig.~\ref{fig:eps} are referenced to $E_{\mathrm{BCS}}^{(0)}$ by running a calculation with no impurity attached.

\section{QP truncation errors}
\label{truncation}
We assess QP-truncation errors by varying the cutoff $Q$ independently of
the bath quadrature with \texttt{qdjj\_solver}~\cite{BCSQPSolver2026}.
For $U/\Delta=2$, $\Gamma/\Delta=0.4$, $D/\Delta=100$, and
$\phi=3\pi/5$ at $W=0$, Table~\ref{tab:qp_cutoff_convergence} shows the
variational energies for $Q=1,2,3,4,5,6$ on a common grid of 32 signed levels per
bath. The energies decrease monotonically as the variational space is
enlarged. Repeating the calculation at each cutoff with 24 levels gives a maximum energy
change of $1.7\times10^{-10}\Delta$: bath discretization errors are
much smaller than QP truncation errors.
The scalar level-3 approximation in Eq.~\eqref{eq:2QP_app2} adds a further $8.9 
\times 10^{-4}\Delta$ energy error within the 2-QP
space.

The DMRG reference calculation uses 24 signed normal-state levels per
reservoir and a maximum matrix-product-state bond dimension of 192.
Increasing the number of levels from 24 to 32 at bond dimension 128
changes the energy by less than $2\times10^{-10}\Delta$;
increasing the bond dimension from 128 to 192 at 24 levels changes
it by less than $3\times10^{-10}\Delta$.
We obtain $\epsilon_{\mathrm{ref}}/\Delta=-1.08852071$ and adopt
a conservative numerical uncertainty of $10^{-8}\Delta$.
The 6-QP result agrees with it within this uncertainty.

\begin{table}[!htbp]
\caption{Convergence with the QP-number cutoff for $W=0$.
The last column gives the discrepancy from the DMRG result in the final
row.}
\label{tab:qp_cutoff_convergence}
\centering
\begin{tabular}{crr}
\hline\hline
$Q$ & $\epsilon/\Delta$ & $(\epsilon-\epsilon_{\mathrm{ref}})/\Delta$ \\
\hline
1 & $-0.9701028968$ & $1.18\times10^{-1}$ \\
2 & $-1.0776041901$ & $1.09\times10^{-2}$ \\
3 & $-1.0879935942$ & $5.27\times10^{-4}$ \\
4 & $-1.0885072004$ & $1.35\times10^{-5}$ \\
5 & $-1.0885205280$ & $1.82\times10^{-7}$ \\
6 & $-1.0885207082$ & $\text{unresolved}$ \\
DMRG & $-1.08852071(1)$ & --- \\
\hline\hline
\end{tabular}
\end{table}

Convergence was also checked for the empty-dot probability at $\phi=0$,
$D/\Delta=100$, and $U/\Delta=0,2,4$. At $\Gamma=\Delta$, the numerical
2-QP projection (without kernel truncation) underestimates the DMRG probability by $7.9$--$10.1\%$.
Increasing the cutoff to $Q=3$ reduces this underestimate to
$0.98$--$1.67\%$, and $Q=4$ reduces it below $0.12\%$. The full
hybridization dependence and comparisons with the scalar closures are
shown in Fig.~{S2} of the Supplemental Material.

The full linear spin response $R_Q$ in Fig.~\ref{fig:sector_contributions}(d)
provides a more sensitive test. At the parameters of
Table~\ref{tab:qp_cutoff_convergence}, it
differs from its $Q=6$ value by $20.8\%$,
$1.77\%$, and $0.052\%$ for $Q=2,3,4$, respectively.
Table~{S3} of the Supplemental Material gives
the cutoff sequence and its bath-refinement check. These results
illustrate why convergence must be assessed separately for each observable.

The $Q=3$ coupling benchmarks in Fig.~\ref{fig:EJES} use $U/\Delta=2$,
$D/\Delta=100$, and $\Gamma/\Delta=0.1,0.4,1$.
The conventional coupling is evaluated at $W=0$, while the spin splitting
is calculated at $\phi=\pi/2$ with both background amplitudes $W_N$ and $W_S$
included in the Hamiltonian, matching the DMRG reference.
These calculations use 40 signed bath levels per lead; repeating them
with 32 levels changes either coupling by less than $2\times10^{-9}$ of
its DMRG value. Bath-refinement effects are therefore much smaller than
the deviations shown in Fig.~\ref{fig:EJES}(c)--(d).

\section{Anomalous behavior at \texorpdfstring{$U=-2\Delta$}{U=-2 Delta}}
\label{appX}

\newcommand{\peff}{p_{\mathrm{eff}}}
In our previous work, we identified a crossover region around $U=2\Delta$ between Andreev-like and Yu--Shiba--Rusinov-like singlet states in the even-fermion-parity sector~\cite{ilicin2025}. As a diagnostic tool, we used the effective local scaling exponent of the energy as a function of $\Gamma$:
\begin{equation}
    \peff = \frac{\dd \ln\!\left(|\epsilon(\Gamma) - \epsilon(0)|/\Delta\right)}{\dd \ln(\Gamma/\Delta)} \>.
\end{equation}
\begin{figure}[!htbp]
    \centering
    \includegraphics{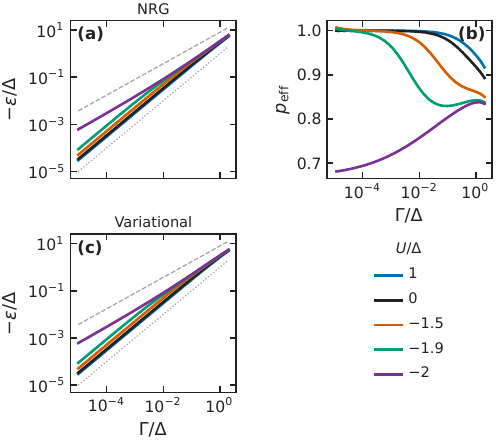}
    \caption{Anomalous behavior around $U=-2\Delta$ at $\phi=0$. (a) NRG binding energies. For $U > -2\Delta$, the
    behavior is linear for small $\Gamma$ (parallel to the dotted line). For $U=-2\Delta$, we find $\epsilon \propto \Gamma^{2/3}$ for small $\Gamma$ (parallel to the dashed line). (b) Effective local scaling exponent $p_\mathrm{eff}$ for the curves in panel (a).
    (c) Variational binding energies from the level-3 approximation, Eq.~\eqref{eq:2QP_app2}, for the same interactions.}
    \label{fig:anomalija}
\end{figure}

As $\Gamma \to 0$, we found $\peff=1$ except at $U = 2 \Delta$, where we observed the anomalous $\sim \Gamma^{2/3}$ behavior. At nonzero hybridization, this point broadens into a crossover region centered on $U / \Delta=2$.
In the odd-parity sector studied in this work, a similar crossover occurs for an attractive interaction around $U=-2\Delta$, where the dominant low-energy weight shifts from the 0-QP component $\ket{\psi_1}$ to the 1-QP sector, represented by $\ket{\psi_2}$ and $\ket{\psi_3}$. The local moment thus moves from the dot orbital to the leads as $U$ becomes more negative.
This picture is supported by the diverging linear-expansion coefficient [Eq.~\eqref{eq:expansion}] at $U=-2\Delta$, as well as by the NRG data presented in Fig.~\ref{fig:anomalija}, where we find $\lim_{\Gamma \to 0} \peff = 2/3$ at the same point.
Although attractive interactions are not relevant for the devices considered here, the singular structure affects a broader parameter range; its signature is visible even for $U \geq 0$ [see Fig.~\ref{fig:anomalija}(b)].
Near the threshold, the gap-edge limit of Eq.~\eqref{eq:1QP} gives $y^2(y+d)\simeq2\Delta\Gamma^2$, where $y=-\epsilon$ and $d=\Delta+U/2$; at $d=0$, this yields $y\propto\Gamma^{2/3}$.
The nonanalytic $\Gamma^{2/3}$ behavior cannot be captured by any finite-order expansion in $\Gamma$.

\section{Wavefunction coefficients in energy space}
\label{appwf}

\begin{figure}[!tb]
    \centering
    \includegraphics{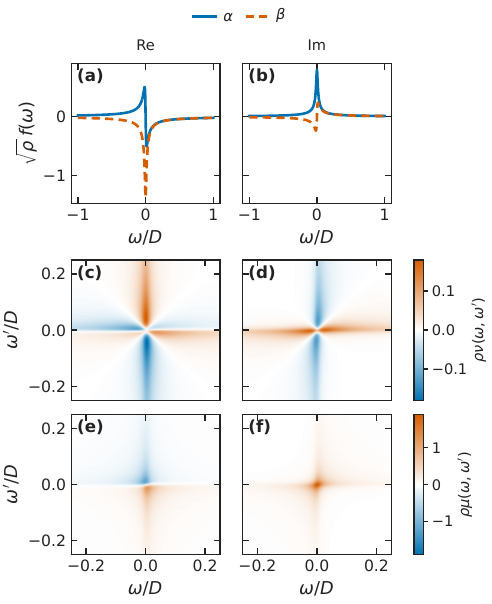}
    \caption{Real (left column) and imaginary (right column) parts of the weighted amplitudes $\sqrt\rho\,\alpha(\omega)$ and $\sqrt\rho\,\beta(\omega)$ [(a), (b)], $\rho\nu(\omega,\omega')$ [(c), (d)], and $\rho\mu(\omega,\omega')$ [(e), (f)], as functions of the normal-state energies $\omega$ and $\omega'$. Computed using the level-3 approximation at $\Gamma / \Delta = 0.4$, $U/\Delta = 2$, and $\phi = 3\pi/5$, with $a>0$ and the phase convention of Eq.~\eqref{eq:coherence_combinations}. In the pair maps, the horizontal coordinate $\omega$ labels the $+$ channel and the vertical coordinate $\omega'$ the $-$ channel. Blue, white, and vermilion denote negative, zero, and positive values; each pair-amplitude row has a shared symmetric color scale covering both parts. Integrating the squared moduli of these weighted amplitudes over $\dd\omega$ or $\dd\omega\,\dd\omega'$ gives the component weights in Eq.~\eqref{eq:component_weights}.}
    \label{fig:wf_omega}
\end{figure}

Figure~\ref{fig:wf_omega} shows the wavefunction coefficients for the reference parameters
$\Gamma / \Delta = 0.4$, $U/\Delta = 2$, and $\phi = 3\pi/5$.

The 1-QP amplitudes in Fig.~\ref{fig:wf_omega} are concentrated near the superconducting gap
edge ($\omega\simeq0$, $E\simeq\Delta$) and exhibit complementary particle-hole parities inherited
from the $a(\omega)$ and $s(\omega)$ BCS factors: $\operatorname{Re} \alpha$ and $\operatorname{Im} \beta$ are
odd in $\omega$, whereas $\operatorname{Im} \alpha$ and $\operatorname{Re} \beta$ are even.

The singlet maps [panels (c) and (d)] have a pronounced cross-shaped
structure, with enhanced amplitude when one QP lies near the gap edge
and the other has a higher excitation energy. The triplet amplitude [panels (e) and (f)] is much larger and
is concentrated where both QPs lie near the gap edge.

This difference reflects interference between the two orders in which
the QPs are created through virtual charge fluctuations. Both sequences return the
impurity to single occupancy. Their contributions enter with opposite signs in the singlet
channel and with the same sign in the triplet channel. Indeed, to leading order in
hybridization, Eqs.~\eqref{eq:nu} and \eqref{eq:mu} give
\begin{equation}
 \frac{\nu(\omega,\omega')}{\mu(\omega,\omega')}
 \simeq
 \frac{E(\omega)-E(\omega')}
 {\sqrt{3}\,[E(\omega)+E(\omega')+U]},
 \label{eq:pair_amplitude_ratio}
\end{equation}
where $E(\omega)=\sqrt{\omega^2+\Delta^2}$. 
The singlet amplitude is suppressed for $E(\omega)\simeq E(\omega')$, giving greater relative
weight to configurations with unequal QP excitation energies.
For the chosen parameters,
the triplet accounts for approximately $95\%$ of the total 2-QP weight. 
The predominance of the triplet component
shows that the 2-QP admixture mainly involves a bath spin $S=1$ coupled
antiferromagnetically to the impurity spin, giving total spin $1/2$.
This triplet admixture supplies the spin-compensation term in
Eq.~\eqref{eq:impurity_spin}, providing an explicit wavefunction
picture of how screening correlations develop while the total
state remains a spin doublet~\cite{Moca2021Kondo}.

\vspace{\baselineskip}

\end{document}


\title{Supplemental Material for\texorpdfstring{\\}{ }
``Quasiparticle-resolved variational theory of Andreev spin qubits''}

\author{Teodor Ili\v cin}
\affiliation{Jo\v zef Stefan Institute, Jamova 39, SI-1000 Ljubljana, Slovenia}
\affiliation{Faculty of Mathematics and Physics, University of Ljubljana, Jadranska 19, SI-1000 Ljubljana, Slovenia}

\author{Rok \v Zitko}
\affiliation{Jo\v zef Stefan Institute, Jamova 39, SI-1000 Ljubljana, Slovenia}
\affiliation{Faculty of Mathematics and Physics, University of Ljubljana, Jadranska 19, SI-1000 Ljubljana, Slovenia}

\date{October 12, 2026}
\maketitle

\section{Gauge transformation under a \texorpdfstring{$2\pi$}{2pi} phase shift}
\label{sec:sm_phase_shift}

The Hamiltonian is $2\pi$-periodic up to a gauge transformation. In the
gauge of the main text, this transformation exchanges the two QP channels.
This channel exchange explains the longer periodicities of individual variational
amplitudes and conditional orbital profiles, while physical observables
remain $2\pi$-periodic.

\subsection{Gauge equivalence and channel exchange}

The dot--lead hopping amplitudes are
$V_L=(V/\sqrt2)e^{-i\phi/4}$ and
$V_R=(V/\sqrt2)e^{i\phi/4}$. Together with the direct hopping amplitude
$W_\sigma=e^{-i\phi/2}(W_N+i\sigma W_S)$, they satisfy
\begin{equation}
 \begin{aligned}
 V_L(\phi+2\pi)&=-iV_L(\phi),\\
 V_R(\phi+2\pi)&=iV_R(\phi),\\
 W_\sigma(\phi+2\pi)&=-W_\sigma(\phi).
 \end{aligned}
 \label{eq:sm_phase_hoppings}
\end{equation}
Using the dot occupancy operator
$n=\sum_\sigma d_\sigma^\dagger d_\sigma$ of the main text, define the right-lead electron number
$\hat N_R=\sum_{k\sigma}c_{Rk\sigma}^\dagger c_{Rk\sigma}$, and introduce
\begin{equation}
 \mathcal G=e^{i\pi n/2}e^{i\pi\hat N_R}.
 \label{eq:sm_phase_unitary}
\end{equation}
Its action on the electron operators is
\begin{equation}
 \begin{aligned}
 \mathcal Gd_\sigma\mathcal G^\dagger&=-i d_\sigma,\\
 \mathcal Gc_{Lk\sigma}\mathcal G^\dagger&=c_{Lk\sigma},\\
 \mathcal Gc_{Rk\sigma}\mathcal G^\dagger&=-c_{Rk\sigma}.
 \end{aligned}
 \label{eq:sm_phase_electrons}
\end{equation}
Thus $c_L^\dagger d$, $c_R^\dagger d$, and $c_L^\dagger c_R$ acquire
the factors $-i$, $i$, and $-1$, respectively, reproducing
Eq.~\eqref{eq:sm_phase_hoppings}. The lead pairing terms are unchanged
because both right-lead creation operators acquire a minus sign. The
lead kinetic energy and dot Hamiltonian are also invariant. Consequently,
\begin{equation}
 H(\phi+2\pi)=\mathcal G H(\phi)\mathcal G^\dagger,
 \label{eq:sm_phase_hamiltonian}
\end{equation}
so the spectra at the two phase biases coincide.

The Bogoliubov operators have phase-independent coherence factors in
this gauge. Equation~\eqref{eq:sm_phase_electrons} therefore leaves
$\gamma_{Lk\sigma}$ unchanged and changes the sign of
$\gamma_{Rk\sigma}$. Applying this to the channel definition in
Eq.~\textup{(15)} gives
\begin{equation}
 \begin{aligned}
 \mathcal G\gamma_{s,k,\sigma}\mathcal G^\dagger
 &=\frac{\gamma_{L,k,\sigma}-s\gamma_{R,-\bar k,\sigma}}{\sqrt2}\\
 &=\gamma_{-s,k,\sigma}.
 \end{aligned}
 \label{eq:sm_phase_channels}
\end{equation}
The exchange of $+$ and $-$ thus follows from a sign change in one lead.

\subsection{Transformation of the variational amplitudes}

We now specialize to the $W=0$, $\delta=0$ doublet Ansatz in
Eqs.~\textup{(20)}--\textup{(24)}, with the nonzero
0-QP coefficient chosen real and positive, $a>0$. The reference state
$|\Phi_0\rangle$ is invariant under $\mathcal G$: the dot is empty and
each BCS component has even right-lead electron parity. Meanwhile,
\begin{equation}
 \begin{aligned}
 \mathcal Gd_\up^\dagger|\Phi_0\rangle
 &=i d_\up^\dagger|\Phi_0\rangle,\\
 \mathcal G(1\pm d_\up^\dagger d_\dn^\dagger)\mathcal G^\dagger
 &=1\mp d_\up^\dagger d_\dn^\dagger.
 \end{aligned}
 \label{eq:sm_phase_dot_states}
\end{equation}
Combining Eq.~\eqref{eq:sm_phase_dot_states} with
Eq.~\eqref{eq:sm_phase_channels} exchanges the entire configurations
multiplying $\alpha_k$ and $\beta_k$. To retain $a>0$ on the
corresponding doublet branch, we choose
\begin{equation}
 |\Psi(\phi+2\pi)\rangle=-i\mathcal G|\Psi(\phi)\rangle.
 \label{eq:sm_phase_state}
\end{equation}
The scalar coefficient obeys $a(\phi+2\pi)=a(\phi)$, while
\begin{equation}
 \begin{aligned}
 \alpha_k(\phi+2\pi)&=-i\beta_k(\phi),\\
 \beta_k(\phi+2\pi)&=-i\alpha_k(\phi).
 \end{aligned}
 \label{eq:sm_phase_one_qp}
\end{equation}

For a 2-QP configuration, the channel exchange must be followed by
fermionic reordering to restore the convention that the $+$ creation
operator precedes the $-$ operator. This reordering exchanges the two energy
indices. The reordering sign is canceled by the antisymmetric singlet
spin combination, but retained for the triplet. The phase from the
singly occupied dot is canceled by the overall $-i$ in
Eq.~\eqref{eq:sm_phase_state}, giving
\begin{equation}
 \begin{aligned}
 \nu_{kk'}(\phi+2\pi)&=\nu_{k'k}(\phi),\\
 \mu_{kk'}(\phi+2\pi)&=-\mu_{k'k}(\phi).
 \end{aligned}
 \label{eq:sm_phase_two_qp}
\end{equation}
The first and second indices label the $+$ and $-$ channels,
respectively; these relations transpose the orbital coordinates rather
than interchange singlet and triplet sectors. They can also be checked
directly in the variational equations using the coherence combinations
of Eq.~\textup{(30)}:
\begin{equation}
 s_k(\phi+2\pi)=-i a_k(\phi),\qquad
 a_k(\phi+2\pi)=-i s_k(\phi).
 \label{eq:sm_phase_coherence}
\end{equation}

\subsection{Component and observable periodicities}

Together with complex conjugation, which maps $\phi\to-\phi$,
Eqs.~\eqref{eq:sm_phase_one_qp} and~\eqref{eq:sm_phase_two_qp} give the reflection relations
of Eq.~\textup{(35)}. Applying Eq.~\eqref{eq:sm_phase_one_qp} twice gives
\begin{equation}
 \alpha_k(\phi+4\pi)=-\alpha_k(\phi),\qquad
 \beta_k(\phi+4\pi)=-\beta_k(\phi),
 \label{eq:sm_phase_antiperiodicity}
\end{equation}
whereas two applications of Eq.~\eqref{eq:sm_phase_two_qp} undo the transposition and sign, so
$\nu$ and $\mu$ are $4\pi$-periodic. Channel-resolved weights, densities, and natural orbitals
therefore have period $4\pi$ at a fixed channel label.

The summed 1-QP weight $w_1=|\alpha|^2+|\beta|^2$ and the individual
0-QP, singlet-pair, and triplet-pair weights are $2\pi$-periodic.
More generally, a physical observable represented by
$O(\phi+2\pi)=\mathcal G O(\phi)\mathcal G^\dagger$ satisfies
\begin{equation}
 \langle O(\phi+2\pi)\rangle_{\phi+2\pi}
 =\langle O(\phi)\rangle_\phi.
 \label{eq:sm_phase_observables}
\end{equation}
Local charge and spin operators, including the dot--lead spin
correlations, are invariant under $\mathcal G$. Their expectation
values, the doublet energy, and the Josephson current
$I=(2e/\hbar)\partial_\phi\epsilon$ therefore repeat after $2\pi$.
The longer component periodicities reflect gauge-dependent channel
labels and phase conventions, rather than a fractional $4\pi$
Josephson effect.

\section{Background Josephson energy}
\label{sec:EJbg}

To obtain the leading contribution to $E_J$ through the direct interlead hopping, set $\Gamma=0$
and apply second-order perturbation theory in $H_T$. Pair creation
and annihilation give a 2-QP intermediate state of energy
$E(\omega)+E(\omega')$. Summing over both spins and inserting
the negative resolvent gives the phase-dependent shift
\begin{equation}
 \begin{split}
 \delta\epsilon_{T,\phi}^{(2)}={}&-(\rho W)^2\cos\phi\\
 &\times\int_{-D}^{D}\!\dd\omega\int_{-D}^{D}\!\dd\omega'
 \frac{\Delta^2}{E(\omega)E(\omega')[E(\omega)+E(\omega')]}.
 \end{split}
 \label{eq:background_integral}
\end{equation}
The spin-dependent phases cancel between the two electrons of a
singlet pair. With 
\begin{equation}
\omega=\Delta\sinh t, \quad 
\omega'=\Delta\sinh s,
\end{equation}
the wide-band integral is
\[
 4\Delta\int_0^\infty\dd t\int_0^\infty\dd s\,
 \frac{1}{\cosh t+\cosh s}=\pi^2\Delta,
\]
yielding Eq.~\textup{(67)} and reproducing the Ambegaokar--Baratoff result~\cite{ambegoakar1963}.

\section{SOC parameter estimation}
\label{weakES}

The experimentally fitted magnitudes $|E_S|/h$ are $300$ and $560$\,MHz
in Ref.~\onlinecite{bargerbos2023} and $305$\,MHz in
Ref.~\onlinecite{pitavidal2023}; in these references, the coefficient, denoted $E_{\mathrm{SO}}$, multiplies a
Pauli matrix, so the maximum zero-field splitting is $2|E_S|$.
To set a scale, we choose
$\Delta/h=46$\,GHz ($190\,\mu\mathrm{eV}$), as obtained from combined NRG and circuit-model
fitting in a related device~\cite{bargerbos2022}, and use our 
reference parameters $U/\Delta=2$, $\Gamma/\Delta=0.4$. The leading weak-hybridization result, Eq.~\textup{(65)}, then gives
\begin{equation}
 \begin{aligned}
 |\rho W_S|&\simeq\frac{|E_S|}{\pi\Gamma}
       \left(1+\frac{U}{2\Delta}\right)\\
       &=0.011\qquad(|E_S|/h=305\,\mathrm{MHz}).
 \end{aligned}
 \label{eq:mag_experimental_scale}
\end{equation}
This value fixes only the spin-dependent amplitude. Taking $\chi=2\pi/3$, as in
Fig.~{9}, gives
\begin{equation}
 \rho W=\frac{|\rho W_S|}{|\sin(\chi/2)|}\simeq0.012.
 \label{eq:experimental_background_hopping}
\end{equation}
This value lies in the weak-background-tunneling regime. For these parameters,
Eq.~\textup{(69)} gives $r_{\mathrm{bg}}\simeq0.11$.

\section{Finite-band Josephson couplings and two-QP truncation errors}
\label{sec:sm_finite_gamma}
\label{sec:finite_gamma}

We give the leading finite-band coefficients used in main-text
Figs.~{6} and {9}, derive the leading
level-3 kernel error, and explain the bandwidth dependence caused by
undressed 2-QP intermediate states. We then describe the size-consistent
dressing that reduces this artifact. Throughout, $\delta=0$ and the
reference state is the spin-up doublet at $W=0$; the spin response is
evaluated to first order in $W_S$.

\subsection{Leading couplings at finite bandwidth}

Write $u=U/\Delta\geq0$, $g=\Gamma/\Delta$, and
$c_B=\operatorname{arcsinh}(D/\Delta)$. On the positive half of the
band, $\omega=\Delta\sinh t$ gives the dimensionless QP energy
$x=E(\omega)/\Delta=\cosh t$, with $0\leq t\leq c_B$.
Introduce
\begin{equation}
 \langle f\rangle_D=\int_0^{c_B}f(\cosh t)\,\dd t,\qquad
 F_D(z)=\left\langle\frac1{x+z}\right\rangle_D.
 \label{eq:fg_measure}
\end{equation}
For a function $h$ of QP energy, both signs of $\omega$ and the
Jacobian are accounted for by
\[
 \int_{-D}^{D}h(E(\omega)/\Delta)\,\dd\omega
 =2\Delta\langle xh(x)\rangle_D;
\]
the density-of-states factor $\rho$ is included separately.
For $z>-1$, the substitution $q=\tanh(t/2)$, with
$\zeta=\tanh(c_B/2)$, gives~\cite[Eqs.~2.444.1--2]{gradshteyn2007}
\begin{equation}
 F_D(z)=\begin{cases}
 \displaystyle\frac{2}{\sqrt{1-z^2}}
  \arctan\!\left(\zeta\sqrt{\frac{1-z}{1+z}}\right),&-1<z<1,\\[5pt]
 \zeta,&z=1,\\[3pt]
 \displaystyle\frac{2}{\sqrt{z^2-1}}
  \operatorname{artanh}\!\left(\zeta\sqrt{\frac{z-1}{z+1}}\right),&z>1.
 \end{cases}
 \label{eq:fg_FD}
\end{equation}

The 1-QP charge excitation costs $\Delta(x+u/2)$, while a 2-QP
state with a singly occupied dot costs $\Delta(x+y)$, where
$y=E(\omega')/\Delta$. Integrating over $y$ at fixed $x$ gives
\begin{equation}
 \begin{aligned}
 S_D(x)&=\left\langle\frac1{(x+y)(y+u/2)}\right\rangle_D
       =\frac{F_D(u/2)-F_D(x)}{x-u/2},\\
 R_D(x)&=\left\langle\frac{y}{(x+y)(y+u/2)}\right\rangle_D
       =F_D(x)-\frac u2 S_D(x).
 \end{aligned}
 \label{eq:fg_pair_kernels}
\end{equation}
The apparent pole is removable:
$S_D(u/2)=-F_D'(u/2)$ and
$R_D(u/2)=F_D(u/2)+(u/2)F_D'(u/2)$, where the
prime denotes differentiation with respect to the argument.

Performing one integration in main-text
Eqs.~\textup{(53)} and \textup{(64)}, with the finite
upper QP-energy limit $\sqrt{1+(D/\Delta)^2}$, yields
\begin{equation}
 \begin{aligned}
 \mathcal J_D(u)&=\frac1{\pi^2}\left\langle
  \frac{2F_D(x)}{(x+u/2)^2}+\frac{S_D(x)}{x+u/2}\right\rangle_D,\\
 \mathcal S_D(u)&=\frac{F_D(u/2)^2}{\pi}
  +\frac2\pi\left\langle\frac{F_D(x)}{x+u/2}\right\rangle_D.
 \end{aligned}
 \label{eq:fg_energy_moments}
\end{equation}
These coefficients retain the finite bandwidth exactly at leading
order and determine the leading-order curves in main-text
Figs.~{6} and {9} through
\begin{equation}
 \begin{aligned}
 E_{J,D}^{(\Gamma^2)}&=\frac{\Gamma^2}{\Delta}\mathcal J_D(u),\\
 E_{S,D}^{(\rho W\Gamma)}&=2\rho W_S\Gamma\,\mathcal S_D(u).
 \end{aligned}
 \label{eq:fg_leading_couplings}
\end{equation}
The two terms in $\mathcal S_D$ are the $11$ and $02$
contributions, respectively. At these leading orders the energy
depends on phase as $E_{J,D}^{(\Gamma^2)}\cos\phi$ and
$\sigma E_{S,D}^{(\rho W\Gamma)}\sin\phi$.

For $D\gg U,\Delta$, the coefficients reduce to the main-text
closed forms. With $F(z)=\lim_{D/\Delta\to\infty}F_D(z)$ from
Eq.~\textup{(54)}, the inner integral obeys
$F(\cosh t)=t/\sinh t$. The remaining convolution is
\begin{equation}
 \begin{aligned}
 T(u)&=\int_0^\infty\frac{t\,\dd t}{\sinh t(\cosh t+u/2)}\\
     &=\frac{\pi^2}{2(2+u)}-\frac{F(u/2)^2}{2}.
 \end{aligned}
 \label{eq:fg_convolution}
\end{equation}
To establish this identity, integration by parts gives
\[
 (1-u^2/4)T(u)=\int_0^\infty
 \ln\frac{\cosh t+u/2}{\sinh t}\,\dd t-\frac{u\pi^2}{8}.
\]
The logarithmic integral has derivative $F(u/2)/2$ with respect to $u$
and value $\pi^2/8$ at $u=0$, so it equals
$\pi^2/4-(1-u^2/4)F(u/2)^2/2$.
Partial fractions and differentiation then give
\begin{equation}
 \begin{aligned}
 \pi\mathcal S(u)&=F(u/2)^2+2T(u)=\frac{\pi^2}{2+u},\\
 \pi^2\mathcal J(u)&=-4T'(u)+\frac{2T(u)-F(u/2)^2}{u}.
 \end{aligned}
 \label{eq:fg_leading_wide_band}
\end{equation}
The second term in the Josephson expression is the mixed integral
evaluated in Ref.~\onlinecite[Eqs.~(47)--(48)]{novotny2005}.
Using $F'(z)=[zF(z)-1]/(1-z^2)$ at $z=u/2$ reproduces main-text
Eq.~\textup{(55)}; the first line gives
$\mathcal S(u)=\pi/(2+u)$.
Apparent singularities at $u=0,2$ are understood by continuity,
giving the limiting values quoted in the main text.

\subsection{Leading error of the level-3 kernel closure}
\label{sec:sm_closure_error}

Eliminating $\nu,\mu$ leaves coupled integral equations for $\alpha$
and $\beta$. The mixed terms describe a transition from one of these
1-QP configurations to the other through a virtual 2-QP state.
The level-3 approximation keeps one substitution of this coupling,
but omits the subsequent return to the original 1-QP configuration.
Each mixed-kernel substitution supplies one power of $g$, so the
omitted pair of substitutions first changes the energy at order
$\Gamma^3/\Delta^2$.
The leading difference between the full 2-QP first cosine harmonic
and the scalar result $E_J^{(3)}$ is
\begin{align}
 E_{J,\mathrm{2QP}}-E_J^{(3)}
 &=\frac{\Gamma^3}{\pi^3\Delta^2}X_D(u)
 +O(\Gamma^4/\Delta^3),\nonumber\\
 X_D(u)&=\left\langle\frac{2R_DS_D-xS_D^2}{x+u/2}\right\rangle_D.
 \label{eq:fg_closure_deficit}
\end{align}
The wide-band integral has the closed form
\begin{align}
 X(u)&=\frac{2\pi^2[(u+4)F(u/2)-6]}{(u-2)^2(u+2)},\nonumber\\
 X(0)&=\frac{\pi^2(\pi-3)}2,\qquad X(2)=\frac{\pi^2}{60}.
 \label{eq:fg_X_closed}
\end{align}
The function $X_D$ quantifies the leading kernel-truncation error
examined in main-text Fig.~{7}.
The 3-QP contribution to the full-model coefficient occurs at the
same order and is outside this calculation.

\subsection{Bandwidth dependence from undressed two-QP states}
\label{sec:sm_bandwidth_artifact}

The leading doublet energy shift is
\begin{equation}
 \begin{aligned}
 \frac{\epsilon}{\Delta}&=-\frac g\pi G_D+O(g^2),\\
 G_D&=2c_B-uF_D(u/2).
 \end{aligned}
 \label{eq:fg_reference_shift}
\end{equation}
Here $G_D$ is the finite-band counterpart of $G(U)$ in main-text
Eq.~\textup{(41)}. Writing $s=x+y$, the 2-QP denominator
$N=\epsilon-E(\omega)-E(\omega')$ therefore obeys
\begin{equation}
 \begin{aligned}
 \frac N\Delta&=-s-\frac g\pi G_D+O(g^2),\\
 \frac1N&=-\frac1{\Delta s}
  \left[1-\frac{gG_D}{\pi s}+O(g^2)\right].
 \end{aligned}
 \label{eq:fg_bare_pair_denominator}
\end{equation}
The expansion is taken at fixed bandwidth. The explicit band logarithm
$c_B\simeq\ln(2D/\Delta)$ enters the relative correction to the
pair amplitudes, producing bandwidth-dependent terms of order
$\Gamma^3/\Delta^2$ in $E_J$ and $(\rho W_S)\Gamma^2/\Delta$ in $E_S$.

The reference doublet is lowered by charge fluctuations, whereas
a 2-QP configuration retains a singly occupied dot whose own charge
fluctuations require a third QP. Their omission leaves the pair
denominators too large in magnitude and suppresses the couplings.
In the full model these intermediate states acquire a comparable
shift. At $U=0$, the Hamiltonian is quadratic and the exact couplings
have a bandwidth-independent wide-band limit at fixed $\Gamma$ and
$\rho W$.

\subsection{Size-consistent dressing of the 2-QP denominators}
\label{sec:sm_dressing}

We dress the 2-QP denominators to cancel the leading band-logarithmic
mismatch identified in previous subsection. This improves
$E_J$ and the retained $11+02$ spin response, but does not remove the
bandwidth dependence of the complete spin-response operator.
We replace the 2-QP denominator of main-text
Eq.~\textup{(32)} with
\begin{equation}
 \begin{aligned}
 \widetilde N(\epsilon,\omega,\omega')
  &=N(\epsilon,\omega,\omega')\\
  &\quad-\Sigma_1\bigl(\epsilon-E(\omega)-E(\omega')\bigr),\\
 \Sigma_1(z)&=\frac{\Gamma}{\pi}\int_{-D}^{D}
  \frac{\dd\omega''}{z-E(\omega'')-U/2}\\
  &=-\frac{\Gamma}{\pi}I_D(U/2-z),
 \end{aligned}
 \label{eq:sm_dressed_denominator}
\end{equation}
where the finite-band 1-QP integral is
\begin{equation}
 \begin{aligned}
 I_D(h)&=\int_{-D}^D\frac{\dd\omega}{E(\omega)+h}\\
       &=2c_B-2\frac h\Delta F_D(h/\Delta),\qquad h>-\Delta,
 \end{aligned}
 \label{eq:fg_one_qp_integral}
\end{equation}
with $F_D$ defined in Eq.~\eqref{eq:fg_FD}.
The function $\Sigma_1$ is the 1-QP self-energy of the 1-QP closure,
$\epsilon^{(1)}=\Sigma_1(\epsilon^{(1)})$ [main-text
Eq.~\textup{(36)}], evaluated at the energy of the 2-QP configuration.
For a singly occupied dot, the squared amplitudes of all hops that
create a third QP in a given mode sum to $|V_L|^2+|V_R|^2$ for either
dot spin, because the coherence factors obey $u_k^2+v_k^2=1$
for each mode $k$.
The shift is therefore the same for the singlet ($\nu$) and triplet
($\mu$) families.
Equation~\eqref{eq:sm_dressed_denominator} is the exact Schur complement
of a 3-QP block in which each 2-QP configuration couples to its own
copy of the dot-fluctuation continuum.
It neglects Pauli blocking between the third QP and the existing two, and
processes in which the third QP recombines with one of the two existing QPs.
These connected 3-QP terms contain no band logarithm.
The dressed closure energy is not guaranteed to be an upper bound to the
lowest energy in the same symmetry sector. The variational bound
established in main-text Eq.~\textup{(49)} applies to the
Rayleigh quotient evaluated with the physical Hamiltonian.

At leading order, $\epsilon\simeq\Sigma_1(0)=-(\Gamma/\pi)I_D(U/2)$, and
\begin{equation}
 \begin{aligned}
 \frac{\widetilde N}{\Delta}&\simeq -s-\frac{2g}{\pi}\\
 &\quad\times\bigl[(u/2+s)F_D(u/2+s)-(u/2)F_D(u/2)\bigr].
 \end{aligned}
 \label{eq:sm_dressed_leading}
\end{equation}
Compared with Eq.~\eqref{eq:fg_bare_pair_denominator}, the terms
$2c_B$ of $\epsilon$ and $\Sigma_1$ cancel. At fixed $u,s$, the
remaining bracket has a finite wide-band limit.

In the level-3 closure, $N(\epsilon,\omega,\omega')$ is replaced by
$\widetilde N$ in main-text Eqs.~\textup{(33)}--\textup{(45)},
in $A^{(3)}$ and $B^{(3)}$, and in the reconstruction of $\nu$ and $\mu$.
The root of the closure is found as before.
The implicit 3-QP components carry the weight
\begin{equation}
 w_3=-\rho^2\!\iint\!\dd\omega\,\dd\omega'
 \bigl[|\nu|^2+|\mu|^2\bigr]\,
 \Sigma_1'\bigl(\epsilon-E(\omega)-E(\omega')\bigr),
 \label{eq:sm_dressed_weight}
\end{equation}
which is added to the normalization of main-text
Eq.~\textup{(47)}.
The empty-dot probability becomes $P_0=(|\alpha|^2+|\beta|^2+w_3)/2$.
For the full dressed 2-QP eigenproblem, the same expression follows
from the Hellmann--Feynman derivative $P_0=\partial\epsilon/\partial U$.
The spin response uses the $11$ and $02$ terms of main-text
Eq.~\textup{(59)} with the normalized dressed state.

Table~\ref{tab:sm_dressing_dmrg} compares both closures with the DMRG
couplings of Fig.~\ref{fig:sm_dmrg_couplings} and with the numerical
$Q=3$ projection of Sec.~\ref{sec:sm_numerical_benchmarks}.
Throughout $0\leq U/\Delta\leq4$, the dressing reduces the deviations of
$E_J$ by factors of $2.2$--$3.8$ and those of $E_S$ by factors of
$1.5$--$2.8$.
At $U/\Delta=2$ and $\Gamma/\Delta=1$, the energy error at $\phi=0$
decreases from $0.077\Delta$ to $0.032\Delta$, and that of $P_0$ from
$-9.5\%$ to $-4.2\%$.
The remaining discrepancies include contributions from connected 3-QP
processes, scalar-kernel closure, and omitted higher-QP sectors and
spin-operator terms. The numerical $Q=3$ projection captures the connected
3-QP processes and substantially improves the agreement with DMRG.

\begin{table}[!htb]
 \caption{Relative deviations (\%) from DMRG of the couplings in main-text
 Fig.~{9} at $U/\Delta=2$ and $D/\Delta=100$.
 The conventional coupling is evaluated at $W=0$; the spin comparison
 uses $\phi=\pi/2$ and $W/\Gamma=0.4$.
 Level~3 is the undressed closure used in the main text;
 ``dressed'' uses Eq.~\eqref{eq:sm_dressed_denominator}.
 The closures use the $11+02$ response for $E_S$, while the numerical
 $Q=3$ calculation uses $E_S=|\epsilon_\up-\epsilon_\dn|/2$.
 Closures use 48 signed sinh-quadrature nodes; 64 nodes change all entries
 by less than $10^{-11}$ in relative terms.}
 \label{tab:sm_dressing_dmrg}
 \centering
 \begin{tabular}{c|rrr|rrr}
 \hline\hline
 & \multicolumn{3}{c|}{$E_J$} & \multicolumn{3}{c}{$E_S$}\\
 $\Gamma/\Delta$ & level 3 & dressed & $Q=3$ & level 3 & dressed & $Q=3$\\
 \hline
 0.1 & $-7.8$ & $-2.2$ & $-0.1$ & $-5.2$ & $-2.2$ & $-0.1$\\
 0.4 & $-25.9$ & $-9.2$ & $-1.7$ & $-18.0$ & $-8.6$ & $-1.8$\\
 1 & $-43.6$ & $-19.5$ & $-6.9$ & $-30.1$ & $-16.8$ & $-7.6$\\
 \hline\hline
 \end{tabular}
\end{table}

At $U=0$, where the model is quadratic, the bandwidth dependence provides
a direct test.
The exact couplings, obtained from a Bogoliubov--de Gennes diagonalization of the
discretized model, change by less than $1\%$ between $D/\Delta=10^2$ and
$10^4$.
Table~\ref{tab:sm_dressing_bandwidth} shows that the undressed closure
inherits the $\ln(D/\Delta)$ dependence of Sec.~\ref{sec:sm_bandwidth_artifact}.
Dressing substantially reduces the bandwidth dependence over the range shown.
The same test separates the operator blocks.
After dressing, at $\Gamma/\Delta=0.4$ and $U=0$, the $11$ and $02$
terms change from $0.129$ and $0.476$ to $-0.030$ and $0.635$
(in units of $\rho W_S\Delta$) between $D/\Delta=10^2$ and $10^4$, while
their sum changes by less than $0.3\%$.
The $22$ term, evaluated in the same state, changes from $-0.043$ to
$-0.098$; its compensating contributions involve 3-QP configurations.
This motivates retaining the $11+02$ response in the dressed comparison;
it is not the complete next-order spin response.
The individual sector shares at finite $\Gamma$ depend on the bandwidth,
whereas the leading-order shares of main-text
Eq.~\textup{(72)} do not.

\begin{table}[!htb]
 \caption{Couplings at $U=0$ relative to the exact Bogoliubov--de Gennes values at the same
 bandwidth. $E_S$ is the $11+02$ response at $\phi=\pi/2$.}
 \label{tab:sm_dressing_bandwidth}
 \centering
 \begin{tabular}{cc|rr|rr}
 \hline\hline
 & & \multicolumn{2}{c|}{$E_J/E_J^{\mathrm{exact}}$}
 & \multicolumn{2}{c}{$E_S/E_S^{\mathrm{exact}}$}\\
 $\Gamma/\Delta$ & $D/\Delta$ & level 3 & dressed & level 3 & dressed\\
 \hline
 0.4 & $10^2$ & 0.724 & 0.896 & 0.815 & 0.894\\
 0.4 & $10^3$ & 0.635 & 0.885 & 0.770 & 0.889\\
 0.4 & $10^4$ & 0.565 & 0.878 & 0.733 & 0.886\\
 1 & $10^2$ & 0.581 & 0.813 & 0.740 & 0.830\\
 1 & $10^3$ & 0.468 & 0.786 & 0.684 & 0.821\\
 1 & $10^4$ & 0.389 & 0.768 & 0.640 & 0.815\\
 \hline\hline
 \end{tabular}
\end{table}

\section{Relation to the cotunneling treatment}
\label{app:cotunneling}
\newcommand{\hg}{\Lambda}

We relate our results to the cotunneling approach of Padurariu and Nazarov (PN)~\cite{padurariu2010}.
We consider the contribution of the active orbital $0$ and an additional empty orbital $M$ with energy $\hg$. We define $A=E_L+U/2$, $B=E_R+U/2$, and $C=E_L+E_R$ with $E_{L/R}=\sqrt{\xi_{L/R}^2+\Delta^2}$. We set $\xi_0=0$ and take the large-$\Lambda$ limit, so that $\Lambda \gg E_L, E_R, U, \Delta$. Here PN's charging energies are $E(\pm e)=U/2$. The limit is taken first at fixed bandwidth, followed by the wide-band limit $D\gg U,\Delta$; the effective-model parameters are understood after downfolding, as detailed below. The kernels $J_{ee}$ and $J_{eh}$ that enter Eqs.~(22) and (23) of Ref.~\onlinecite{padurariu2010}, evaluated at $(\xi_M,\xi_0)=(\Lambda,0)$, take the form
\begin{equation}
    \lim_{\Lambda \to \infty} \Lambda J_{ee} = \frac{1}{AB} + \frac{1}{AC} + \frac{1}{BC} = \frac{A+B+C}{ABC} \equiv K.
    \label{eq:PN_K}
\end{equation}
The same expression is found for $J_{eh}$. We write
\begin{equation}
    \lim_{\Lambda \to \infty} \Lambda (J_{ee}+J_{eh}) = 2K.
    \label{eq:PN_kernel_limit}
\end{equation} Setting $E_L=\Delta x$, $E_R=\Delta y$, and $u=U/\Delta$ gives
\begin{equation}
    \Delta^2 K = \frac{(x+u/2)+(y+u/2)+(x+y)}{(x+u/2)(y+u/2)(x+y)},
\end{equation}
which has the same form as Eq.~\textup{(64)}. To track the integration measure, we use $\xi\,\dd\xi=E\,\dd E$ on each half of the band. Thus, including both signs of $\xi$ in each lead,
\begin{equation}
 \frac{\Delta^2}{4}
 \int_{-\infty}^{\infty} \frac{\dd\xi_L}{E_L}
 \int_{-\infty}^{\infty} \frac{\dd\xi_R}{E_R}
 K = \pi \mathcal{S}(u).
 \label{eq:PN_measure}
\end{equation}
To determine the normalization factors, we write $T_{ij}$ for the spin matrix multiplying $c_i^\dag c_j$, where $c_i$ and $c_j$ encode bath ($L$/$R$) or dot ($0$, $M$) orbitals; the factors $1/\sqrt N$ are kept implicit. With $v_l=|V_l|$, $t_l>0$, and $\vartheta=\chi/2$, a choice consistent with the gauge of the main text is
\begin{equation}
\begin{aligned}
 T_{L0} &= v_L  e^{-i\phi/4} \mathbf1,&
 T_{R0} &= v_R  e^{i\phi/4}  \mathbf1,\\
 T_{LM} &= t_L  e^{-i\phi/4} \mathbf1,&
 T_{RM} &= -t_R e^{i\phi/4}  e^{-i\vartheta\sigma_z},
\end{aligned}
\label{eq:PN_hoppings}
\end{equation}
with $T_{ji}=T_{ij}^\dag$. The relative minus sign in $T_{RM}$ is chosen so that the empty-orbital denominator $-1/\Lambda$ produces a positive effective hopping magnitude:
to leading order in $1/\Lambda$, elimination of $M$ gives
\begin{equation}
 -\frac{T_{LM} T_{MR}}{\Lambda}
 =W e^{-i\phi/2}e^{i\vartheta\sigma_z},
 \qquad W=\frac{t_L t_R}{\Lambda},
 \label{eq:PN_downfold}
\end{equation}
whose diagonal entries are exactly the amplitudes $W_\sigma$ in Eq.~\textup{(11)}.

With our convention that $T_{ij}$ enters as $c_i^\dag T_{ij}c_j$, PN's definition is $\mathsf P_{M0}=\rho^2(T_{MR}T_{R0}T_{0L}T_{LM})^{\mathsf T}$. The superconducting phases cancel in this product. The transpose implements PN's spin-index order and is immaterial for the diagonal matrices used here. Writing $p=v_L v_R t_L t_R$, we find
\begin{equation}
\begin{split}
  \left(T_{MR} T_{R0} T_{0L} T_{LM}\right)^{\mathsf T}
  &= -p e^{i\vartheta\sigma_z}\\
  &= -p\cos\vartheta\,\mathbf1
    -ip\sin\vartheta\,\sigma_z.
\end{split}
\label{eq:PN_ordered_product}
\end{equation}
For constant $\rho$ per spin and per lead, the two spectral sums give $\rho^2$. Decomposing $\mathsf P_{M0}$ into its scalar and vector parts gives
\begin{equation}
\begin{aligned}
 \mathsf P_{M0} &\equiv P_{M0}^0\mathbf1
       +i\mathbf P_{M0} \cdot \boldsymbol{\sigma}
       = -\rho^2 p e^{i\vartheta\sigma_z},\\
 P_{M0}^0 &= -\rho^2 p \cos\vartheta, \qquad
 \mathbf P_{M0} = -\rho^2 p \sin\vartheta \, \hat{\mathbf z}.
\end{aligned}
\label{eq:PN_P_matrix}
\end{equation}
The reversed orbital ordering gives $\mathsf P_{0M}=\mathsf P_{M0}^\dag$;
hence $P_{0M}^0=P_{M0}^0$ and $\mathbf P_{0M}=-\mathbf P_{M0}$.
No spin trace is taken in these definitions. 
Since $\rho$ is per spin and per lead,
$v_L v_R=|V|^2/2=\Gamma/(2\pi\rho)$. With $W=t_Lt_R/\Lambda$ from Eq.~\eqref{eq:PN_downfold}, the pseudovector becomes
\begin{equation}
    \frac{\mathbf P_{M0}}{\Lambda}
    = -\rho^2 |V_L V_R| W \sin(\chi/2)\,\hat{\mathbf z}
    = -\frac{\rho W \Gamma}{2\pi}\sin(\chi/2)\,\hat{\mathbf z}.
    \label{eq:PN_P_mapping}
\end{equation}

To write the energy denominators explicitly for the six allowed tunneling sequences in each
kernel, we use the charging term $U(n_0+n_M-1)^2/2$.
For fixed lead-electron energies, let $A_\Lambda=A+\Lambda$, $B_\Lambda=B+\Lambda$, $F=\Lambda+2U$, and $G=\Lambda+C$, with $A,B,C$ defined above. PN's six-term kernels can then be grouped as
\begin{align}
 J_{ee} &= \frac{1}{C} \left( \frac{1}{A_\Lambda B} + \frac{1}{AB_\Lambda} \right)
 \nonumber\\
 &\quad + \frac{1}{F} \left( \frac{1}{A_\Lambda} + \frac{1}{A} \right)
                 \left( \frac{1}{B_\Lambda} + \frac{1}{B} \right),
 \label{eq:PN_Jee_explicit}\\
 J_{eh} & =\frac{1}{C} \left( \frac{1}{A_\Lambda A} + \frac{1}{B_\Lambda B} \right)
 \nonumber\\
 &\quad + \frac{1}{G} \left( \frac{1}{A_\Lambda} + \frac{1}{B} \right)
                 \left( \frac{1}{B_\Lambda} + \frac{1}{A} \right).
 \label{eq:PN_Jeh_explicit}
\end{align}
Each kernel has the arguments $(E_L, E_R, \Lambda, 0)$, with the lead QP energies written explicitly here.
These expressions show that $\Lambda J_{ee} \to K$ and $\Lambda J_{eh} \to K$ in the $\Lambda \to \infty$ limit.

The fourth-order amplitude is obtained from 
$\mathcal{V} \mathcal{R} \mathcal{V} \mathcal{R} \mathcal{V} \mathcal{R} \mathcal{V}$, 
where $\mathcal{V}$ is the microscopic dot--lead tunneling, $H_0$ is the unperturbed two-orbital Hamiltonian, and $E_0$ is its ground-doublet energy. The reduced resolvent $\mathcal{R}=Q_\perp(E_0-H_0)^{-1}Q_\perp$ uses the projector $Q_\perp$ onto the complement of that doublet.
For transfer from $L$ to $R$ through both orbitals, each assignment of the two transferred spins has six nonzero time orderings. We label the hops $1:L\to0$, $2:0\to R$, $3:L\to M$, and $4:M\to R$. The chronological strings and their three positive intermediate excitation energies are
\begin{center}
\begin{tabular}{clcl}
$ee$ & energies & $eh$ & energies \\
\hline
1234 & $A,C,B_\Lambda$ & 2134 & $B,C,B_\Lambda$ \\
1324 & $A,F,B_\Lambda$ & 2314 & $B,G,B_\Lambda$ \\
1342 & $A,F,B$ & 2341 & $B,G,A$ \\
3124 & $A_\Lambda,F,B_\Lambda$ & 3214 & $A_\Lambda,G,B_\Lambda$ \\
3142 & $A_\Lambda,F,B$ & 3241 & $A_\Lambda,G,A$ \\
3412 & $A_\Lambda,C,B$ & 3421 & $A_\Lambda,C,A$
\end{tabular}
\end{center}
For initial spin $\sigma$, the transferred spin on orbital $0$ is $-\sigma$ in $ee$ and $\sigma$ in $eh$, with the opposite spin on $M$. The relative fermionic signs are $+$ and $-$, respectively. Combining the three negative resolvents with the minus sign of the hopping product gives Eqs.~\eqref{eq:PN_Jee_explicit} and \eqref{eq:PN_Jeh_explicit}. Omitting the common forward phase $e^{i\phi}$, the directed amplitude is
\begin{equation}
\begin{split}
 \mathcal A_\sigma^{L\to R}
 &= C_{LR} \left[ J_{ee} e^{-i\sigma\vartheta}
                 -J_{eh} e^{i\sigma\vartheta} \right],\\
 C_{LR} &= \frac{\Delta^2p}{4E_LE_R}, \qquad \sigma=\pm1.
\end{split}
\label{eq:PN_directed_amplitude}
\end{equation}
Here $\Delta^2/(4E_L E_R)$ is the product of the anomalous coherence factors $\Delta/(2E_L)$ and $\Delta/(2E_R)$.
Adding the reverse process yields
\begin{equation}
\begin{split}
 \delta\epsilon_\sigma
 &= 2\operatorname{Re}\left[\mathcal{A}_\sigma^{L\to R} e^{i\phi} \right] \\
 &=      2C_{LR} (J_{ee}-J_{eh}) \cos\vartheta \cos\phi \\
 &\quad+ 2C_{LR} (J_{ee}+J_{eh}) \sigma \sin\vartheta \sin\phi.
\end{split}
\label{eq:PN_energy_from_amplitude}
\end{equation}
The signed coefficient of $\sigma\sin\phi$ is $-2\operatorname{Im}\mathcal A_\sigma^{L\to R}/\sigma$. Since $P_{M0,z}=-\rho^2p\sin\vartheta$, in continuum notation it is
\begin{equation}
 E_S^{(0M)}=-2\,\frac{\Delta^2}{4}
 \int\frac{\dd\xi_L}{E_L}\int\frac{\dd\xi_R}{E_R}
 P_{M0,z}(J_{ee}+J_{eh}).
\label{eq:PN_spin_coefficient}
\end{equation}
The factor $2$ in Eq.~\eqref{eq:PN_spin_coefficient} comes from
adding forward and reverse amplitudes, as in PN's Eq.~(19).
For comparison, when only the selected empty orbital $M$ and
the singly occupied orbital $0$ are retained, PN's Eq.~(23) reads
\begin{equation}
\begin{split}
 \boldsymbol{\varepsilon}_{\mathrm{so},M0}^{\mathrm{PN}}
 &= \int_{-\infty}^{\infty}\dd\xi_L
    \int_{-\infty}^{\infty}\dd\xi_R\,q_{LR}\\
 &\quad\times\mathbf P_{M0}(J_{ee}+J_{eh}),
 \qquad q_{LR}=\frac{\Delta^2}{4E_LE_R}.
\end{split}
\label{eq:PN23}
\end{equation}
Here the kernels are PN's $J_{ee,eh}(\xi_L,\xi_R,\Lambda,0)$,
and $\mathbf P_{M0}$ already includes $\rho^2$.
The prefactor in PN's Eq.~(23) is $q_{LR}$, while the magnitude
of the prefactor in Eq.~\eqref{eq:PN_spin_coefficient} is $2q_{LR}$.
PN's Eq.~(1) contains
$(\boldsymbol{\varepsilon}_{\mathrm{so}}\cdot\boldsymbol{\sigma})
\sin\phi$; thus we compare their Pauli-operator coefficient with
our coefficient $E_S^{(0M)}$ of $\sigma\sin\phi$.
Their full level splitting is
$2|\boldsymbol{\varepsilon}_{\mathrm{so}}||\sin\phi|$.
For this collinear contribution,
$|E_S^{(0M)}|
=2|\boldsymbol{\varepsilon}_{\mathrm{so},M0}^{\mathrm{PN}}|$.

Combining Eq.~\eqref{eq:PN_spin_coefficient} with $J_{ee}+J_{eh}\to2K/\Lambda$, Eq.~\eqref{eq:PN_measure}, and Eq.~\eqref{eq:PN_P_mapping} gives
\begin{equation}
\begin{split}
    E_S &= -4\pi \mathcal{S}(u) \frac{P_{M0,z}}{\Lambda} \\
        &= 2\rho W\Gamma \sin(\chi/2) \mathcal{S}(u).
\end{split}
\label{eq:PN_ES_match}
\end{equation}

\section{Relation to the atomic-limit treatment}
\label{app:atomic}

To see how Eq.~\textup{(66)} matches the scaling found in Ref.~\onlinecite{bargerbos2023}, we first note that $U/\Delta \to 0$ in the superconducting atomic limit, so that $\mathcal{S}(0)=(2/\pi)\bigl[\int_1^\infty \dd x/(x\sqrt{x^2-1})\bigr]^2 = \pi/2$.
The spin-conserving and spin-dependent active-orbital hopping amplitudes used by Bargerbos et al.\ are $v$ and $w_\mathrm{sf}$, respectively; we define $T_L = v+iw_\mathrm{sf}\sigma_x$ and $T_R = v-iw_\mathrm{sf}\sigma_x$. We introduce $\tau=\sqrt{v^2+w_\mathrm{sf}^2}$ and $\tan\theta=w_\mathrm{sf}/v$, so that $T_L=\tau e^{i\theta \sigma_x}$ and $T_R=\tau e^{-i\theta\sigma_x}$.

Independent lead spin rotations make both hoppings equal to $\tau$ and transform the direct hopping into $t e^{-2i\theta\sigma_x}$. We then use a global rotation $R=e^{i\pi\sigma_y/4}$ satisfying $R^\dag\sigma_xR=-\sigma_z$, so the direct hopping becomes $t e^{2i\theta\sigma_z}$. This maps to our convention as $W=t$ and $\chi/2=2\theta$.

Our $\Gamma$ is the total two-lead hybridization, whereas Bargerbos et al.\ use per-lead quantities; hence $\Gamma=2(\Gamma_V+\Gamma_W)$ with $\Gamma_V=\pi \rho v^2$ and $\Gamma_W=\pi \rho w_\mathrm{sf}^2$. For positive hopping magnitudes $v,w_\mathrm{sf},t$, $\sin(2\theta)=2vw_\mathrm{sf}/(v^2+w_\mathrm{sf}^2)=2\sqrt{\Gamma_V \Gamma_W}/(\Gamma_V+\Gamma_W)$.

Then Eq.~\textup{(66)} becomes
$E_S=\pi \rho t \Gamma \sin(2\theta) = 4 \pi \rho t \sqrt{\Gamma_V \Gamma_W}$. Using $\tilde t=\pi \rho t$, we find
\begin{equation}
    E_S=4\tilde{t} \sqrt{\Gamma_V \Gamma_W}.
\end{equation}
This expression equals the small-$\tilde t$, superconducting-atomic-limit coefficient in Eq.~(S24) of the Supplemental Material of Ref.~\onlinecite{bargerbos2023}. Under the specified rotation, their term $-E_{\mathrm{SO}}\sigma_x\sin\phi$ in Eq.~(S25) of that reference becomes $+E_S\sigma_z\sin\phi$.

\section{Tight-binding representation of a flat-DOS band}
\label{sec:sm_real_space}

\subsection{Canonical transformation and numerical reconstruction}

To obtain site-resolved observables, we transform the
continuum amplitudes to the Legendre basis.
We normalize the continuum modes by
\[
 \{c_{l\sigma}(\omega),c^\dag_{l'\sigma'}(\omega')\}
 =\delta_{ll'}\delta_{\sigma\sigma'}\delta(\omega-\omega')/\rho.
\]
In terms of the orthonormal Legendre functions $p_n(\omega)$ defined in
Eq.~\textup{(82)}, the orthogonal transformation~\cite{chin2010,devega2015}
is
\begin{equation}
 c_{ln\sigma}=\int_{-D}^{D}\rho\,\dd\omega\,
 p_n(\omega)c_{l\sigma}(\omega).
 \label{eq:rs_canonical}
\end{equation}
Orthogonality gives canonical site anticommutators, and the Legendre
recurrence relation gives the chain hopping $t_n$ in Eq.~\textup{(83)}.
We choose real standing-wave
modes, so time reversal pairs opposite spins in the same spatial
orbital. Orthogonality then maps
$\int\rho\,\dd\omega\,c^\dag_{l\up}(\omega)c^\dag_{l\dn}(\omega)$
to $\sum_nc^\dag_{ln\up}c^\dag_{ln\dn}$ without altering $\Delta$.
Applying the same transformation to QPs defines $g_{ln\sigma}$. Since
$p_n(-\omega)=(-1)^np_n(\omega)$, the inverse channel rotation is
\begin{equation}
 g_{Ln\sigma}=\frac{g_{+n\sigma}+g_{-n\sigma}}{\sqrt2},\qquad
 g_{Rn\sigma}=(-1)^n\frac{g_{+n\sigma}-g_{-n\sigma}}{\sqrt2}.
 \label{eq:rs_lead_inverse}
\end{equation}

For quadrature weights $w_i$ integrating $\rho\,\dd\omega$, the amplitudes
in the normalized discrete basis form vectors $\mathbf A,\mathbf B$
and matrices $\mathsf F,\mathsf G$:
\begin{equation}
 \begin{aligned}
 A_i&=\sqrt{w_i}\alpha_i,\qquad B_i=\sqrt{w_i}\beta_i,\\
 (\mathsf F_{ij},\mathsf G_{ij})&=\sqrt{w_iw_j}(\nu_{ij},\mu_{ij}).
 \end{aligned}
 \label{eq:rs_weighted_blocks}
\end{equation}
Here $f_i=f(\omega_i)$ and $f_{ij}=f(\omega_i,\omega_j)$ denote sampled
energy-space functions. The integrated component weights of Eq.~\textup{(48)} are
represented by $|\alpha|^2=\|\mathbf A\|^2$, $|\beta|^2=\|\mathbf B\|^2$,
$|\nu|^2=\|\mathsf F\|_F^2$, and $|\mu|^2=\|\mathsf G\|_F^2$;
the squared Frobenius norms sum the squared moduli of all matrix entries.
We use the total 1-QP probability $w_1$ of Sec.~\ref{sec:sm_phase_shift}
and use the main-text total 2-QP probability $w_2$.
We solve the equations on a coarse quadrature grid and extend the solution to the dense target grid described below by evaluating the full integral equations at the target energies. With the
coherence-factor combinations and energy denominators of Eqs.~\textup{(30)} and~\textup{(32)}, the coarse-grid
sums give
\begin{align}
 N_s(\omega)&=N(\omega)
       -V^2\sum_i\frac{w_i|s_i|^2}{N(\omega,\omega_i)},\nonumber\\
 \alpha(\omega)&=\frac{a(\omega)}{N_s(\omega)}
 \left[\frac{Va}{\sqrt2}+
 \frac{V^2}{2}\sum_i\frac{w_i s_i^*\beta_i}{N(\omega,\omega_i)}\right].
 \label{eq:rs_extension}
\end{align}
For $\beta(\omega)$, interchange $a(\omega)\leftrightarrow s(\omega)$,
$s_i\leftrightarrow a_i$, and $\beta_i\leftrightarrow\alpha_i$;
the scalar vacuum amplitude $a$ is unchanged. We reconstruct the pair
coefficients on the target grid from
\begin{equation}
 \begin{aligned}
 \nu(\omega,\omega')&=\frac{V[-s(\omega)\alpha(\omega')+\beta(\omega)a(\omega')]}{2N(\omega,\omega')},\\
 \mu(\omega,\omega')&=-\frac{\sqrt3V[s(\omega)\alpha(\omega')+\beta(\omega)a(\omega')]}{2N(\omega,\omega')}.
 \end{aligned}
 \label{eq:sm_pair_reconstruction}
\end{equation}
This is a Nystr\"om extension
of the full solution.

For an $M$-point Gauss--Legendre target quadrature rule, define
\begin{equation}
    T_{ni}=\sqrt{w_i}p_n(\omega_i),
\end{equation}
with $0\leq n<M$.
Then $TT^{\mathsf T}=I$, and the chain coefficients are
\begin{equation}
 \begin{aligned}
 \boldsymbol\alpha&=T\mathbf A,\qquad \boldsymbol\beta=T\mathbf B,\\
 \boldsymbol\nu&=T\mathsf F T^{\mathsf T},\qquad
 \boldsymbol\mu=T\mathsf G T^{\mathsf T}.
 \end{aligned}
 \label{eq:sm_chain_blocks}
\end{equation}
We use 48 coarse nodes, 1600 target nodes,
and 800 retained sites. Increasing the coarse grid to 64 nodes changes
the normalized state by less than $2\times10^{-14}$ at three test points.
At the reference point $U/\Delta=2$, $\Gamma/\Delta=0.4$,
$\phi=3\pi/5$, and $D/\Delta=100$, increasing the number of target nodes from 1200 to 1600 changes
the transformed amplitude blocks by less than $3\times10^{-9}$ in norm
and the site profiles by less than $5\times10^{-14}$ pointwise.
At the reference point, retaining 800 sites gives a spin-sum-rule residual of less than $4\times10^{-9}$.

\begin{figure}[!tb]
 \centering
 \includegraphics{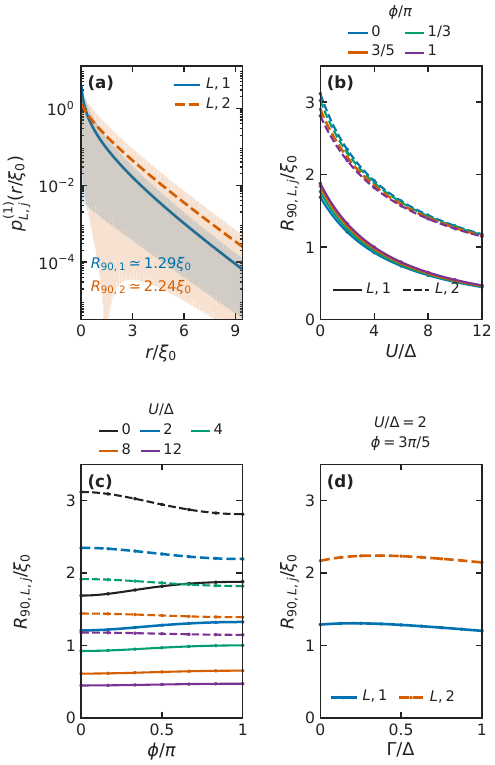}
 \caption{One-QP natural orbitals in the left lead at particle-hole
 symmetry, $W=0$, and $D/\Delta=100$.
 (a) Unit-normalized densities of modes $1$ (solid) and $2$ (dashed),
 ordered by decreasing occupation, at $U/\Delta=2$,
 $\Gamma/\Delta=0.4$, and $\phi=3\pi/5$.
 (b)--(d) $R_{90}$ versus $U/\Delta$, $\phi/\pi$, and
 $\Gamma/\Delta$, respectively; fixed parameters take their reference
 values unless indicated. Colors label modes in (a) and (d) and parameter
 cuts in (b) and (c).
 Adjacent-site probabilities are combined into cells of width
 $h=2\pi\Delta/D$ in $r/\xi_0$, with $\xi_0=Da_0/(\pi\Delta)$,
 and divided by $h$; faint lines show raw-site densities.
 Radii are continuous $90\%$ quantiles of the monotone two-site-cell
 cumulative probability with full-space normalization.
}
 \label{fig:sm_lead_orbitals}
\end{figure}

\subsection{One-QP natural orbitals in a physical lead}
\label{sec:sm_lead_orbitals}

At particle-hole symmetry, Eq.~\eqref{eq:rs_lead_inverse} rotates the
symmetry channels into physical leads. Projecting the 1-QP component
onto $L$ and tracing out the dot and QP spin gives
$(\boldsymbol\alpha\boldsymbol\alpha^\dagger+
\boldsymbol\beta\boldsymbol\beta^\dagger)/2$; the orthogonal dot-charge
factors remove cross terms. Its trace is $w_1/2$, so
$P(L\mid\mathrm{1QP})=1/2$. Conditioning on this projection gives the
unit-trace spatial density matrix
\begin{equation}
 (\varrho_L^{(1)})_{nm}
 =\frac{\alpha_n\alpha_m^*+\beta_n\beta_m^*}{w_1}.
 \label{eq:sm_lead_one_qp_density}
\end{equation}

Its two nonzero eigenvalues are the conditional occupation probabilities
of the left-lead natural orbitals, with $\lambda_1+\lambda_2=1$.
At the reference point, $\lambda_1=0.74318$ and $\lambda_2=0.25682$:
the orbital state is mixed after tracing out the dot, with a preference
for the first mode. Equal eigenvalues $\lambda_1=\lambda_2=1/2$ give
maximal mixing and maximal occupation fluctuations within this two-mode
space. For the mode-occupation operator $\hat n_{L,j}$, the conditional
variance is $\lambda_j(1-\lambda_j)$, approximately $0.191$ here compared
with the maximum $1/4$.
Figure~\ref{fig:sm_lead_orbitals} shows the unit-normalized mode densities
and their containment radii. The right-lead orbitals differ by the
site-parity factor $(-1)^n$, giving identical densities, occupations,
and radii.

\subsection{Conditional QP orbitals}
\label{sec:sm_localization}

Let $\Pi_{\mathrm{2QP}}$ project
onto exactly two bath QPs and write
$|\Psi_{\mathrm{pair}}\rangle=\Pi_{\mathrm{2QP}}|\Psi\rangle/\sqrt{w_2}$.
For a fixed channel $s=\pm1$, the orbital density matrix is
\begin{equation}
 (\varrho_s)_{nm}=\sum_\sigma
 \langle\Psi_{\mathrm{pair}}|g^\dagger_{s,m,\sigma}g_{s,n,\sigma}|\Psi_{\mathrm{pair}}\rangle.
 \label{eq:sm_channel_elements}
\end{equation}
This matrix is the orbital reduced density matrix obtained by tracing out the
dot, both QP spins, and the other channel's orbital coordinate.

In terms of the chain coefficients,
\begin{equation}
 \begin{aligned}
 (\varrho_+)_{nm}&=\frac1{w_2}\sum_j
 (\nu_{nj}\nu_{mj}^*+\mu_{nj}\mu_{mj}^*),\\
 (\varrho_-)_{nm}&=\frac1{w_2}\sum_j
 (\nu_{jn}\nu_{jm}^*+\mu_{jn}\mu_{jm}^*).
 \end{aligned}
 \label{eq:sm_channel_trace}
\end{equation}
The sum over $j$ discards the other QP's position, while the spin trace
removes singlet--triplet cross terms. The spin-summed one-body matrix retaining
\emph{both} channel labels is $\varrho_+\oplus\varrho_-$ and has trace
two, the conditional QP number.

The full channel-resolved position probability is the diagonal of the
reduced density matrix,
$(\varrho_s)_{nn}=\sum_j\lambda_{s,j}|\psi_{s,j}(n)|^2$,
including contributions from all natural orbitals.
Figure~{11}(a) instead shows the probability density of the leading orbital,
normalized to unit probability and expressed as a function of $r/\xi_0$.
The orbital state is generally mixed after tracing out other degrees of freedom. 
Since the dot and spins are also traced
out, the two channel spectra need not coincide even though
$|\Psi_{\mathrm{pair}}\rangle$ is pure.

For example, a factorized spatial kernel
$\boldsymbol\mu=\sqrt{w_2}\mathbf u\mathbf v^{\mathsf T}$,
with $\boldsymbol\nu=0$ and normalized $\mathbf u,\mathbf v$, gives
$\varrho_+=\mathbf u\mathbf u^\dagger$ and
$\varrho_-=\mathbf v\mathbf v^\dagger$. Each has one nonzero eigenvalue, equal to one: the state
contains two well-defined orbital wavepackets, even though its spins
can remain entangled.

\subsection{Physical spin contractions}

To evaluate the spin distribution $q_l(n)$ and the unconnected
dot--lead correlation $C_l(n)$ defined in the main text, we express
the physical electron operators in terms of the normalized discrete
QP operators $\gamma_{s,i,\sigma}$. Here $s=\pm$ labels the symmetry
channel, $i=1,\ldots,M$ labels the nodes of the $M$-point target
quadrature, and $\sigma=\pm1$ denotes spin up or down, with
$\bar\sigma=-\sigma$.

The transformation to chain site $n=0,\ldots,M-1$ is specified
by $T_{ni}=\sqrt{w_i}p_n(\omega_i)$. In the calculations described
above, $M=1600$ and we retain the first 800 sites, $n=0,\ldots,799$.
Combining this transformation with the Bogoliubov transformation
and the channel rotation gives
\begin{equation}
 \begin{aligned}
 c_{Ln\sigma}
 &=\frac{1}{\sqrt2}\sum_{i=1}^{M}T_{ni}
 \Bigl[u_i\bigl(\gamma_{+,i,\sigma}+\gamma_{-,i,\sigma}\bigr)\\
 &\qquad
 +\sigma v_i\bigl(\gamma^\dagger_{+,i,\bar\sigma}
                 +\gamma^\dagger_{-,i,\bar\sigma}\bigr)\Bigr],\\
 c_{Rn\sigma}
 &=\frac{(-1)^n}{\sqrt2}\sum_{i=1}^{M}T_{ni}
 \Bigl[v_i\bigl(\gamma_{+,i,\sigma}-\gamma_{-,i,\sigma}\bigr)\\
 &\qquad
 +\sigma u_i\bigl(\gamma^\dagger_{+,i,\bar\sigma}
                 -\gamma^\dagger_{-,i,\bar\sigma}\bigr)\Bigr].
 \end{aligned}
 \label{eq:rs_electron_rows}
\end{equation}
Here $u_i=u(\omega_i)$ and $v_i=v(\omega_i)$ are the BCS coherence
factors. The right-lead expression includes the energy reflection
implicit in the channel definition: $u(-\omega)=v(\omega)$ and
$p_n(-\omega)=(-1)^np_n(\omega)$.

The spin contractions below use the weighted amplitudes
$\mathbf A$, $\mathbf B$, $\mathsf F$, and $\mathsf G$ of
Eq.~\eqref{eq:rs_weighted_blocks}. To write these contractions
compactly, at each fixed site $n$ we collect the coefficients
appearing in the electron operators into two real
$1\times M$ row vectors:
\[
 \begin{aligned}
 x&=\bigl(T_{n1}u_1,\;T_{n2}u_2,\;\ldots,\;T_{nM}u_M\bigr),\\
 y&=\bigl(T_{n1}v_1,\;T_{n2}v_2,\;\ldots,\;T_{nM}v_M\bigr).
 \end{aligned}
\]
Both vectors depend on the chosen site $n$, which we suppress
in the notation. Multiplication by a weighted
amplitude column, for example, gives
\[
 \begin{aligned}
 x\mathbf A&=\sum_{i=1}^{M}T_{ni}u_i A_i,\\
 y\mathbf A&=\sum_{i=1}^{M}T_{ni}v_i A_i.
 \end{aligned}
\]
Direct contraction gives, for either lead,
\begin{align}
 q_{11}&=\frac12\sum_{r=x,y}(|r\mathbf A|^2+|r\mathbf B|^2),\nonumber\\
 q_{\mu\mu}&=\frac13\sum_{r=x,y}
 (\|r\mathsf G\|^2+\|r\mathsf G^{\mathsf T}\|^2),\nonumber\\
 q_{\nu\mu}&=\frac1{\sqrt3}\operatorname{Re}\sum_{r=x,y}
 \left[\langle r\mathsf G^{\mathsf T},r\mathsf F^{\mathsf T}\rangle
       -\langle r\mathsf G,r\mathsf F\rangle\right],\nonumber\\
 q_{02}&=-\sqrt{\frac23}\operatorname{Re}
 \left[a^*x(\mathsf G-\mathsf G^{\mathsf T})y^{\mathsf T}\right],
 \label{eq:rs_spin_parts}
\end{align}
where $\langle v,w\rangle=\sum_i v_i^*w_i$ and the site argument is
suppressed. Consequently,
\begin{equation}
 \begin{aligned}
 q_l(n)&=q_{11}+q_{\mu\mu}+q_{\nu\mu}+q_{02},\\
 C_l(n)&=\frac34(-q_{\mu\mu}+q_{\nu\mu}+q_{02}).
 \end{aligned}
 \label{eq:rs_spin_contractions}
\end{equation}
The latter expression follows from the Clebsch--Gordan coefficients
and $\langle\mathbf S_d\cdot\mathbf s_l\rangle
=3\langle S_d^zs_l^z\rangle$ in a total-spin-$1/2$ state.
Completeness gives $\sum_nq_{11}=w_1/2$,
$\sum_nq_{\nu\mu}=\sum_nq_{02}=0$, and
$\sum_nq_{\mu\mu}=2|\mu|^2/3$, reproducing the spin and correlation
sum rules of the main text.

\section{Additional numerical benchmarks}
\label{sec:sm_numerical_benchmarks}

\subsection{QP-cutoff convergence of the empty-dot probability}
\label{sec:sm_p0_convergence}

\begin{figure}[!htb]
 \centering
 \includegraphics{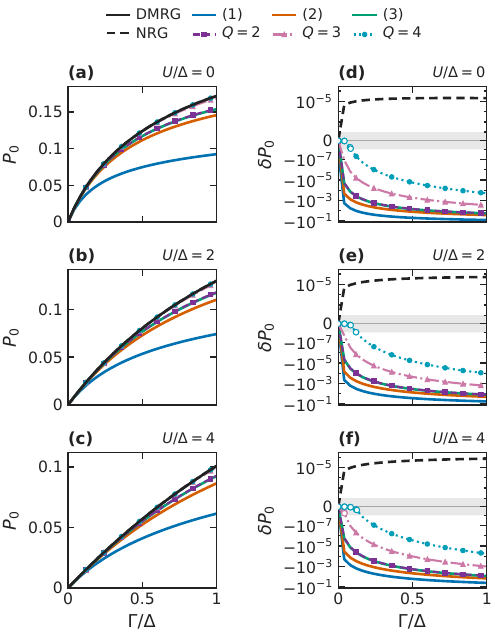}
 \caption{Convergence of the empty-dot probability with the bath-QP cutoff
 at particle-hole symmetry, $\phi=0$, $W=0$, and $D/\Delta=100$.
 Panels (a)--(c) show $P_0$ for $U/\Delta=0,2,4$, respectively;
 panels (d)--(f) show the signed differences
 $\delta P_0=P_0-P_0^{\mathrm{DMRG}}$.
 Numerical $Q=2,3,4$ results are compared with approximation levels
 (1)--(3) and NRG. Approximation level~3 is a scalar kernel closure within the
 2-QP space, whereas numerical $Q=3$ retains 3-QP configurations.
 The discrepancy axes use a symmetric logarithmic scale, linear within
 $\pm10^{-8}$. Shading denotes the empirical DMRG probability uncertainty;
 open circles mark numerical-QP differences within the combined numerical
 resolution. Differences are evaluated at directly matched hybridization
 values.}
 \label{fig:sm_p0_cutoff}
\end{figure}

Figure~\ref{fig:sm_p0_cutoff} compares the empty-dot probability with
DMRG as the maximum number $Q$ of bath QPs is increased. Each projected
Hamiltonian is solved numerically, with bath refinement controlled
independently of $Q$.

We evaluate the physical projector
$P_0=1-\langle n_d\rangle+\langle n_{d\up}n_{d\dn}\rangle$.
At particle-hole symmetry, for the cutoffs $Q=2,3,4$, this also gives
$P_0=(w_1+w_3)/2$, where $w_n$ is the probability of the $n$-QP sector.
Thus the 3-QP charge configurations contribute to $P_0$ when they
are retained.
All three cutoffs are evaluated at the same 50 hybridization
values.
The empirical probability uncertainty from bath refinement is below
$2.7\times10^{-9}$ for the numerical QP results. The DMRG reference has
a probability uncertainty of $10^{-8}$.
At $U=0$, the DMRG probabilities agree with finite-band Bogoliubov--de Gennes
calculations to within $2.2\times10^{-10}$.

\subsection{QP-cutoff convergence of the linear spin response}
\label{sec:sm_spin_convergence}

For the normalized numerical projected eigenstate $\ket{\Psi_Q}$ at $W=0$,
define the dimensionless spin-up response
\begin{equation}
 R_Q=\frac{1}{\Delta}
 \left\langle\Psi_Q\left|
 \frac{\partial H}{\partial(\rho W_S)}
 \right|\Psi_Q\right\rangle_{W=0}.
 \label{eq:sm_spin_response}
\end{equation}
To linear order in $W_S$, the signed spin splitting at fixed $\phi$ is
$\epsilon_\up-\epsilon_\dn=2\rho W_S\Delta R_Q$.
Table~\ref{tab:sm_spin_cutoff} gives the response at the parameters of
main-text Table~{I}:
$U/\Delta=2$, $\Gamma/\Delta=0.4$, $D/\Delta=100$, and $\phi=3\pi/5$.
This phase-point response includes all phase harmonics; it is distinct
from the coupling extracted at $\phi=\pi/2$ in main-text
Fig.~{9}(b). All operator blocks allowed by each cutoff
are included.

\begin{table}[!htb]
 \caption{Convergence of the full linear spin response at the reference
 parameters of main-text Table~{I}.
 Results use 32 signed bath levels per reservoir; comparison with 24
 levels changes $R_Q$ by less than $1.6\times10^{-9}$. The final column
 measures the relative difference from the $Q=6$ result.}
 \label{tab:sm_spin_cutoff}
 \centering
 \begin{tabular}{crr}
 \hline\hline
 $Q$ & $R_Q$ & $R_Q/R_6-1$ \\
 \hline
 2 & $0.36942498$ & $-20.76\%$ \\
 3 & $0.45795936$ & $-1.774\%$ \\
 4 & $0.46598733$ & $-0.05235\%$ \\
 5 & $0.46622873$ & $-0.000576\%$ \\
 6 & $0.46623142$ & --- \\
 \hline\hline
 \end{tabular}
\end{table}

The same calculation quantifies the omitted $22$ contribution.
In the full numerical 2-QP eigenstate at this reference point,
\[
 R_{11}+R_{02}=0.38405038,\qquad R_{22}=-0.01462540,
\]
where off-diagonal contributions include their Hermitian conjugates.
Including $22$ therefore reduces the $11+02$ response by $3.81\%$.

\begin{figure}[!htb]
 \centering
 \includegraphics{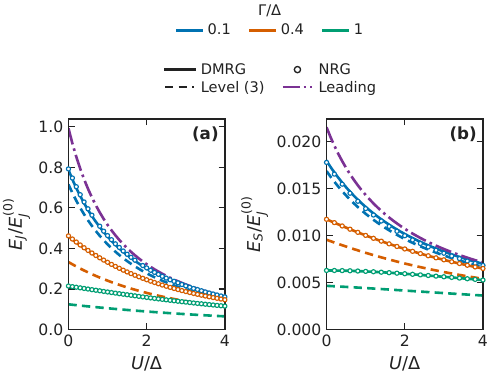}
 \caption{Independent numerical reference for the couplings in main-text
 Fig.~{9}: (a) $E_J=[\epsilon(0)-\epsilon(\pi)]/2$ at
 $W=0$ and (b) $E_S=|\epsilon_\up-\epsilon_\dn|/2$ at $\phi=\pi/2$.
 Solid: DMRG; open circles: NRG; dashed: level~3, using $11+02$ in (b);
 purple dash-dotted: leading finite-band expressions. The largest empirical
 relative DMRG uncertainty in $E_S$ is below $0.039\%$.}
 \label{fig:sm_dmrg_couplings}
\end{figure}

\subsection{Independent DMRG reference for the Josephson couplings}
\label{sec:sm_dmrg_couplings}

Figure~\ref{fig:sm_dmrg_couplings} provides the independent DMRG
cross-check summarized in Sec.~{IV\,E}.

\renewcommand{\bibsection}{\section*{\refname}}